# Large Language Models for Mental Health Applications: A Systematic Review

Zhijun Guo[1]; Alvina Lai[1]; Johan Hilge Thygesen[1]; Joseph Farrington[1]; Thomas Keen[1,2]; Kezhi Li[1]

[1]Institute of Health Informatics, University College London
[2]GOS Institute of Child Health, University College London

**Corresponding author:**
Kezhi Li
Institute of Health Informatics
University College London
222 Euston Road
London
United Kingdom
Phone: +44 7859 995590
Email: ken.li@ucl.ac.uk

## Abstract
**Background**

Large language models (LLMs) are advanced artificial neural networks trained on extensive datasets to accurately understand and generate natural language. While they have received much attention and demonstrated potential in digital health, their application in mental health, particularly in clinical settings, has generated considerable debate.

**Objective:**

This systematic review aims to critically assess the use of LLMs in mental health, specifically focusing on their applicability and efficacy in early screening, digital interventions, and clinical settings. By systematically collating and assessing the evidence from current studies, our work analyzes the models, methodologies, data sources, and outcomes, thereby highlighting the potential of LLMs in mental health, the challenges they present, and the prospects for their clinical use.

**Methods:**

Adhering to the PRISMA guidelines, this review searched five open-access databases: MEDLINE (accessed by PubMed), IEEE Xplore, Scopus, Journal of Medical Internet Research (JMIR), and ACM Digital Library (ACM). Keywords used were: (mental health OR mental illness OR mental disorder OR psychiatry ) AND (large language models). We included articles published between January 1, 2017, and April 30, 2024, and excluded non-English articles.

**Results:**
In total, 40 articles were evaluated, including mental health conditions and suicidal ideation detection through text analysis (n=15), the use of LLMs in mental health conversational agents (CAs) (n=7), and other applications and evaluation of the LLMs in mental health (n=18). LLMs exhibit substantial effectiveness in detecting mental health issues and providing accessible, de-stigmatized eHealth services. However, assessments also indicate that the current risks associated with the clinical use might surpass their benefits. These risks include inconsistencies in generated text, the production of hallucinatory content, and the absence of a comprehensive, benchmarked ethical framework.

**Conclusions:**
This systematic review examines the clinical applications of LLMs in mental health, highlighting their potential and inherent risks. The study identifies several issues: the lack of multilingual datasets annotated by experts, concerns regarding the accuracy and reliability of generated content, challenges in interpretability due to the 'black box' nature of LLMs, and ongoing ethical dilemmas. These ethical concerns include the absence of a clear, benchmarked ethical framework, data privacy issues, and the potential for over-reliance on LLMs by both physicians and patients, which could compromise traditional medical practices. As a result, LLMs should not be considered substitutes for professional mental health services. However, the rapid development of LLMs underscores their potential as valuable clinical aids, emphasizing the need for continued research and development in this area.

**Keywords**
Large language models; Mental health; Digital healthcare; ChatGPT; BERT

___________________________________________________________________________________

## 1. Introduction and Background
### 1.1 Mental Health

Mental health, a critical component of overall well-being, is at the forefront of global health challenges [1]. In 2019, an estimated 970 million individuals worldwide suffered from mental illness, accounting for 12.5% of the global population [2]. Anxiety and depression are among the most prevalent psychological conditions, affecting 301 million and 280 million individuals respectively [2]. Additionally, 40 million people were afflicted with bipolar disorder, 24 million with schizophrenia, and 14 million experienced eating disorders [3]. These mental disorders collectively contribute to an estimated USD 5 trillion in global economic losses annually [4]. Despite the staggering prevalence, many cases remain undetected or untreated, with the resources allocated to the diagnosis and treatment of mental illness far less than the negative impact it has on society [5]. Globally, untreated mental illnesses account for 5% in high-income countries and 19% in low- and middle-income countries [3]. The COVID-19 pandemic has further exacerbated the challenges faced by mental health services worldwide [6], as the demand for these services increased while access was decreased [7]. This escalating crisis underscores the urgent need for more innovative and accessible mental health care approaches.

Mental illness treatment encompasses a range of modalities including medication, psychotherapy, support groups, hospitalization, and complementary & alternative medicine [8]. However, societal stigma attached to mental illnesses often deters people from seeking appropriate care [9]. Many people with mental illness avoid or delay psychotherapy [10], influenced by fears of judgment and concerns over costly, ineffective treatments [11]. The COVID-19 crisis and other global pandemics have underscored the importance of digital tools, such as telemedicine and mobile apps, in delivering care during critical times [12]. In this evolving context, LLMs present new possibilities for enhancing the delivery and effectiveness of mental health care.

Recent technological advancements have revealed some unique advantages of LLMs in mental health. These models, capable of processing and generating text akin to human communication, provide accessible support directly to users [13]. A study analyzing 2,917 Reddit user reviews found that CAs powered by LLMs are valued for their non-judgmental listening and effective problem-solving advice. This aspect is particularly beneficial for socially marginalized individuals, as it enables them to be heard and understood without the need for direct social interaction [14]. Moreover, LLMs enhance the accessibility of mental health services, which are notably undersupplied globally [15]. Recent data reveals substantial delays in traditional mental health care delivery: 23% of individuals with mental illnesses report waiting over 12 weeks for face-to-face psychotherapy sessions [16], with 12% waiting more than six months, and 6% over a year [16]. In addition, 43% of adults with mental illness indicate that such long waits have exacerbated their conditions [15].

Telemedicine, enhanced by LLMs, offers a practical alternative that expedites service delivery and could flatten traditional healthcare hierarchies [17]. This includes real-time counseling sessions through CAs that are not only cost-effective but also accessible anytime and from any location. By reducing the reliance on physical visits to traditional healthcare settings, telemedicine has the potential to decentralize access to medical expertise and diminish the hierarchical structures within the healthcare system [17]. Mental health chatbots developed using language models, have been gaining recognition, such as Woebot [18] and Wysa [19]. Both chatbots follow the principles of Cognitive Behavioural Therapy principles and are designed to equip users with self-help tools for managing their mental health issues [20]. In clinical practice, LLMs hold the potential to support the automatic assessment of therapists' adherence to evidence-based practices and the development of systems that offer real-time feedback and support for patient homework between sessions [21]. These models also have the potential to provide feedback on psychotherapy or peer support sessions, which is especially beneficial for clinicians with less training and experience [21]. Currently, these applications are still in the proposal stage. Although promising, they are not yet widely used in routine clinical settings, and further evaluation of their feasibility and effectiveness is necessary.

The deployment of LLMs in mental health also poses several risks, particularly for vulnerable groups. Challenges such as inconsistencies in the content generated and the production of 'hallucinatory' content may mislead or harm users [22], raising serious ethical concerns. In response, authorities like the World Health Organization (WHO) have developed ethical guidelines for Artificial Intelligence (AI) research in healthcare, emphasizing the importance of data privacy, human oversight, and the principle that AI tools should augment, rather than replace, human practitioners [23]. These potential problems with LLMs in healthcare have gained considerable industry attention, underscoring the need for a comprehensive and responsible evaluation of LLMs' applications in mental health. The following section will further explore the workings of LLMs, and their potential applications in mental health, and critically evaluate the opportunities and challenges they introduce.

### 1.2 Large Language Models

LLMs represent advancements in machine learning (ML), characterized by their ability to understand and generate human-like text with high accuracy [24]. The efficacy of these models is typically evaluated using benchmarks designed to assess their linguistic fidelity and contextual relevance. Common metrics include BLEU for translation accuracy and ROUGE for summarization tasks [25]. LLMs are characterized by their scale, often encompassing billions of parameters, setting them apart from traditional language models [26]. This breakthrough is largely due to the Transformer architecture, a deep neural network structure that employs a 'self-attention' mechanism, developed by Vaswani et al. in 2017. This allows LLMs to process information in parallel rather than sequentially, greatly enhancing speed and contextual understanding [27]. To clearly define the scope of this study concerning LLMs, we specify that an LLM must utilize the Transformer architecture and contain a high number of parameters, traditionally at least one billion, to qualify as 'large' [28]. This criterion encompasses models such as Generative Pre-trained Transformers (GPT) and Bidirectional Encoder Representations from Transformers (BERT). Although the standard BERT model, with only 0.34 billion parameters [29], does not meet the traditional criteria for 'large', its sophisticated bidirectional design and pivotal role in establishing new natural learning processing (NLP) benchmarks justify its inclusion among notable LLMs [30]. The introduction of ChatGPT in 2022 generated substantial public and academic interest in LLMs, underlining their transformative potential within the field of AI [31]. Other state-of-the-art LLMs include LLaMA and PaLM, as illustrated in Figure 1.

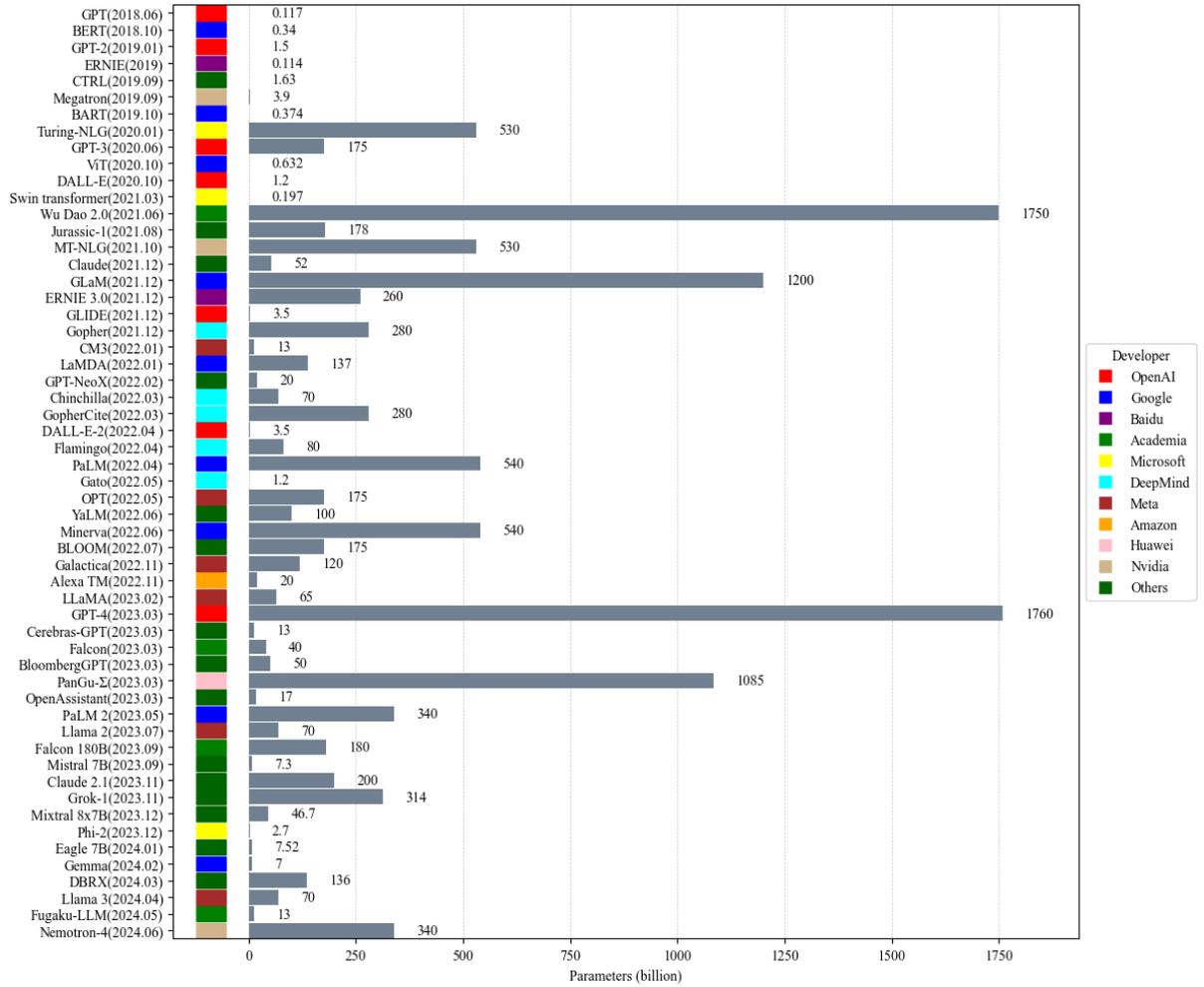

**Figure 1. Comparative analysis of LLMs by parameter size and developer entity.** The bar chart represents the number of parameters in billions for various language models by date of publication, with the oldest models at the top. The legend is color-coded by the development entity. Data was summarized with the latest models up to June 2024, with data for parameters and developers from GPT to LLaMA adapted from the work of Thirunavukarasu AJ et al [32]. Details on release dates, parameter sizes, and developer entities for the most recent LLMs are sourced from references [33-35].

LLMs are primarily designed to learn fundamental statistical patterns of language [36]. Initially, these models were used as the basis for fine-tuning task-specific models rather than training those models from scratch, offering a more resource-efficient approach [37]. This fine-tuning process involves adjusting a pre-trained model to a specific task by further training it on a smaller, task-specific dataset [38]. However, developments in larger and more sophisticated models have reduced the need for extensive fine-tuning in some cases. Notably, some advanced LLMs can now effectively understand and execute tasks specified through natural language prompts without extensive task-specific fine-tuning [39]. Instruction fine-tuned models undergo additional training on pairs of user requests and appropriate responses. This training allows them to generalize across various complex tasks, such as sentiment analysis, which previously required explicit fine-tuning by researchers or developers [40]. A key part of the input to these models, like

ChatGPT and Gemini, includes a system prompt, often hidden from the user, which guides the model on how to interpret and respond to user prompts. For example, it might direct the model to act as a helpful mental health assistant. Additionally, 'prompt engineering' has emerged as a crucial technique in optimizing model performance. Prompt engineering involves crafting input texts that guide the model to produce the desired output without additional training. For example, refining a prompt from 'Tell me about current events in healthcare' to 'Summarize today's top news stories about technology in healthcare' provides the model with more specific guidance, which can enhance the relevance and accuracy of its responses [41]. While prompt engineering can be highly effective and reduce the need to retrain the model, it is important to be wary of 'hallucinations', a phenomenon where models confidently generate incorrect or irrelevant outputs [42]. This can be particularly challenging in high-accuracy scenarios, such as healthcare and medical applications [43-46]. Thus, while prompt engineering reduces the reliance on extensive fine-tuning, it underscores the need for thorough evaluation and testing to ensure the reliability of model outputs in sensitive applications.

The existing literature includes a review of the application of ML and NLP in mental health [47], analyses of LLMs in medicine[32], and a scoping review of LLMs in mental health. These studies have demonstrated the effectiveness of NLP for tasks such as text categorization and sentiment analysis [47] and provided a broad overview of LLM applications in mental health [48]. However, a gap remains in systematically reviewing state-of-the-art LLMs in mental health, particularly in the comprehensive assessment of literature published since the introduction of the Transformer architecture in 2017.

This systematic review addresses these gaps by providing a more in-depth analysis, evaluating the quality and applicability of studies, and exploring ethical challenges specific to LLMs, such as data privacy, interpretability, and clinical integration. Unlike previous reviews, this study excludes preprints, follows a rigorous search strategy with clear inclusion and exclusion criteria (e.g., using Cohen's kappa to assess inter-reviewer agreement), and employs a detailed assessment of study quality and bias (e.g., using the Risk of Bias 2 tool) to ensure the reliability and reproducibility of the findings.

Guided by specific research questions, this systematic review critically assesses the use of LLMs in mental health, focusing on their applicability and efficacy in early screening, digital interventions, and clinical settings, as well as the methodologies and data sources employed. Our findings highlight the potential of LLMs in enhancing mental health diagnostics and interventions, while also identifying key challenges, such as inconsistencies in model outputs and the lack of robust ethical guidelines. These insights suggest that, while LLMs hold promise, their use should be supervised by physicians, and they are not yet ready for widespread clinical implementation.

## 2. Methods

This systematic review followed the Preferred Reporting Items for Systematic Review and Meta-analysis (PRISMA) guidelines [49]. The protocol was registered on PROSPERO

under the ID: CRD42024508617.

## 2.1 Search Strategies

The search was initiated on August 3, 2024, and completed on August 6, 2024, by one author (ZG). This author systematically searched four databases: MEDLINE, IEEE Xplore, Scopus, JMIR, and ACM using the following search keywords: (mental health OR mental illness OR mental disorder OR psychiatry) AND (large language models). These keywords were consistently applied across each database to ensure a uniform search strategy. To conduct a comprehensive and precise search for relevant literature, strategies were tailored for different databases. 'All Metadata' was searched in MEDLINE and IEEE Xplore, while the search in Scopus was confined to titles, abstracts, and keywords. The JMIR database utilized the 'Criteria Exact Match' feature to refine search results and enhance precision. In the ACM database, the search focused on 'Full text'. The screening of all citations involved four steps:

1) Initial Search: All relevant citations were imported into a Zotero citation manager library.

2) Preliminary Inclusion: Citations were initially screened based on predefined inclusion criteria.

3) Duplicate Removal: Citations were consolidated into a single group, from which duplicates were eliminated.

4) Final Inclusion: The remaining references were carefully evaluated against the inclusion criteria to determine their suitability.

## 2.2 Study Selection and Eligibility Criteria

All the articles that matched our search criteria were double-screened by two independent reviewers (ZG, KL) to ensure each article fell within the scope of LLMs in mental health. This process involved the removal of duplicates followed by a detailed manual evaluation of each article to confirm adherence to our predefined inclusion criteria, ensuring a comprehensive and focused review. To quantify the agreement level between the reviewers and ensure objectivity, inter-rater reliability was calculated using Cohen's kappa, with a score of 0.84 indicating a good level of agreement. In instances of disagreement, a third reviewer (AL) was consulted to achieve consensus.

To assess the risk of bias, we utilized the Risk of Bias 2 tool, as recommended for Cochrane Reviews. The results have been visualized in Multimedia Appendix 1. We thoroughly examined each study for potential biases that could impact the validity of the results. These included biases from the randomization process, deviations from intended interventions, missing outcome data, inaccuracies in outcome measurement, and selective reporting of results. This comprehensive assessment ensures the credibility of each study.

The criteria for selecting articles were as follows: We limited our search to English-language publications, focusing on articles published between January 1, 2017, and April 30, 2024. This timeframe was chosen considering the significant developments in the field of LLMs in 2017, marked notably by the introduction of the Transformer architecture, which has greatly influenced academic and public interest in this area.

In this review, the original research articles and available full-text papers have been carefully selected aiming to focus on the application of LLMs in mental health. To comply with PRISMA guidelines, articles that have not been published in a peer-reviewed venue, including those only available on a preprint server, were excluded. Due to the limited literature specifically addressing the mental health applications of LLMs, we included review articles to ensure a comprehensive perspective. Our selection criteria focused on direct applications, expert evaluations, and ethical considerations related to the use of LLMs in mental health contexts, with the goal of providing a thorough analysis of this rapidly developing field.

### 2.3 Information Extraction

The data extraction process was jointly conducted by two reviewers (ZG, KL), focusing on examining the application scenarios, model architecture, data sources, methodologies used, and main outcomes from selected studies on LLMs in mental health.

Initially, we categorized each study to determine its main objectives and applications. The categorization process was conducted in two steps. First, after reviewing all the included articles, we grouped them into three primary categories: detection of mental health conditions and suicidal ideation through text, LLM usage for mental health CAs, and other applications and evaluation of the LLMs in mental health. In the second step, we performed a more detailed categorization. After a thorough, in-depth reading of each article within these broad categories, we refined the classifications based on the specific goals of the studies. Following this, we summarized the main model architectures of the LLMs used and conducted a thorough examination of data sources, covering both public and private datasets. We noted that some review articles lacked detail on dataset content, and therefore, we focused on providing comprehensive information on public datasets, including their origins and sample sizes. We also investigated the various methods employed across different studies, including data collection strategies and analytical methodologies. We examined their comparative structures and statistical techniques to offer a clear understanding of how these methods are applied in practice.

Finally, we documented the main outcomes of each study, recording significant results and aligning them with relevant performance metrics and evaluation criteria. This included providing quantitative data where applicable to underscore these findings. The synthesis of information was conducted using a narrative approach, where we integrated and compared results across different studies to highlight the efficacy and impact of LLMs on mental health. This narrative synthesis allowed us to highlight the efficacy and impact of LLMs in mental health, providing quantitative data where applicable to underscore these findings. The results of our analysis are presented in three tables, each corresponding to one of the primary categories.

## 3. Results
### 3.1 Strategy and Screening Process

The PRISMA diagram of the systematic screening process can be seen in Figure 2. Our initial search across four academic databases: MEDLINE, IEEE Xplore, Scopus, JMIR, and ACM yielded 14265 papers: 907 from MEDLINE, 102 from IEEE Xplore, 204 from Scopus, 211 from JMIR, and 12,841 from ACM. After duplication, 13,967 unique papers were retained. Subsequent screening is based on predefined inclusion and exclusion criteria, narrowing down the selection to 40 papers included in this review. The reasons for the full-text exclusion of 61 papers can be found in Multimedia Appendix 2.

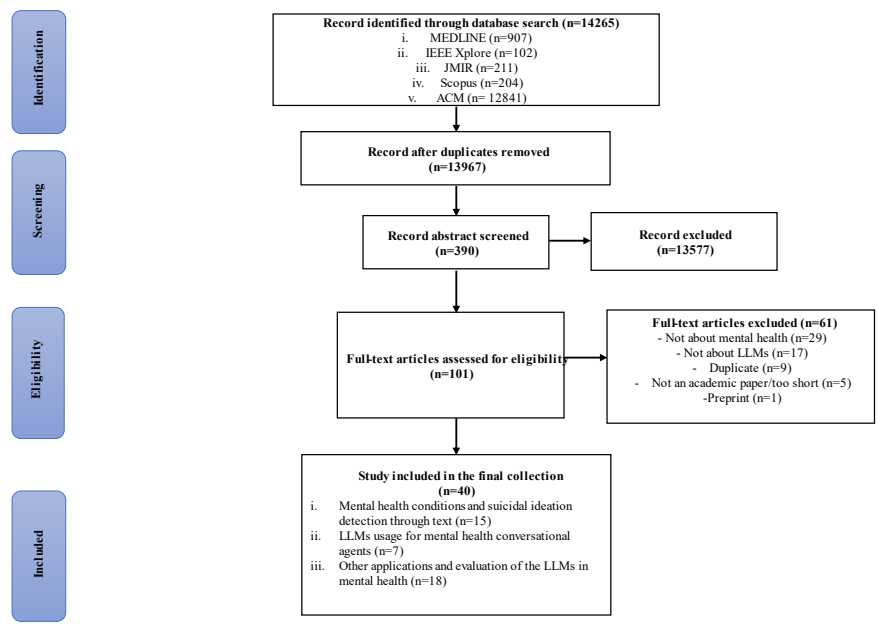

**Figure 2. PRISMA flow of selection process.**

In our review of the literature, we classified the included articles into three broad categories: detection of mental health conditions and suicidal ideation through text (n=15), LLMs usage for mental health CAs (n=7), and the other applications and evaluation of the LLMs in mental health (n=18). The first category investigates the potential of LLMs for the early detection of mental illness and suicidal ideation via social media and other textual sources. Early screening is highlighted as essential for preventing the progression of mental disorders and mitigating more severe outcomes. The second category assesses LLM-supported CAs used as teletherapeutic interventions for mental health issues, such as loneliness, with a focus on evaluating their effectiveness and validity. The third category covers a broader range of LLM applications in mental health, including clinical uses such as decision support and therapy enhancement. It aims to assess the overall effectiveness, utility, and ethical considerations associated with LLMs in these settings. All selected articles are summarized in Table 1-3 according to the three categories.

**Table 1. Summary of the 15 selected articles from the literature on LLMs in mental health conditions and suicidal ideation detection through text.**

| Ref. | Cases | Models | Data Sources | Methodology Used | Main Outcomes |
|---|---|---|---|---|---|
| (Verma et al., 2023) [50] | Detecting depression using LLMs through textual data | RoBERTa [51] | Mental health corpus [52]; Depression Reddit cleaned [53] | This paper used two datasets focused on mental health to train a deep learning model, RoBERTa, for depression detection. Data preprocessing included text cleaning, tokenization, and vectorization, followed by model training with fine-tuning of hyperparameters for binary classification of depression. | The study successfully used a RoBERTa-base model to detect depression with a high accuracy of 96.86%, showcasing the potential of AI in identifying mental health issues through linguistic analysis. |
| (Diniz et al., 2022) [54] | Detecting suicidal ideation using LLMs through Twitter texts | BERT model for Portuguese [55]; Multilingual BERT (base) [29]; BERTimbau [56] | Non-clinical texts from tweets (user posts of the online social network Twitter) | This paper developed the Boamente system through a co-design approach with psychologists, creating a virtual keyboard for text collection and a web platform for data analysis. Texts were collected from Twitter, annotated for suicidal ideation, and processed with LLM techniques. ML and DL models were then trained on this data, allowing mental health professionals to access analysis results without retaining sensitive text data. | The Boamente system demonstrated effective text analysis for suicidal ideation with high privacy standards and actionable insights for mental health professionals. The best-performing BERTimbau Large model (accuracy: 0.955; precision: 0.961; F-score: 0.954; AUC: 0.954) significantly excelled in detecting suicidal tendencies, showcasing robust accuracy and recall in model evaluations. |
| (Danner et al., 2023) [57] | Detecting depression using LLMs through clinical interviews | BERT; GPT-3.5; ChatGPT-4 | DAIC-WOZ [58]; Extended-DAIC [59]; simulated data | The method employed deep learning to detect depression symptoms using multimodal datasets. It pre-processed data, addressed class imbalance, fine-tuned the model, and evaluated performance using metrics like precision, recall, and F1 score. | The study assessed the abilities of GPT-3.5-turbo and ChatGPT-4 on the DAIC-WOZ dataset, which yielded F1 scores of 0.78 and 0.61 respectively, and a custom BERT model, extended-trained on a larger dataset, which achieved an F1 score of 0.82 on the Extended-DAIC dataset, in recognizing depression in text. |
| (Tao et al., 2023) [60] | Detecting anxiety and depression using LLMs through dialogs in real-life scenarios | ChatGPT | Speech data from nine Q&A tasks related to daily activities (75 patients with anxiety and 64 patients with depression) | The study developed a virtual framework utilizing LLMs to non-intrusively support mental health treatments by analyzing behavioral data and speech features like rate and pitch. This approach, tested with patients from Peking University Sixth Hospital, aimed to detect anxiety and depression symptoms through ChatGPT analysis, facilitating personalized therapeutic strategies. | This paper introduced a virtual interaction framework using LLMs to mitigate negative psychological states. Analysis of Q&A dialogues demonstrated ChatGPT's potential in identifying depression and anxiety. To enhance classification, four language features, including prosodic and speech rate, positively impacted classification. |
| (Hayati et al., 2022) [61] | Detecting depression by Malay dialect speech using LLMs | GPT-3 | Interviews with 53 adults fluent in Kuala Lumpur (KL), Pahang, or Terengganu Malay dialects | Participants were interviewed, their responses transcribed and classified for depression using GPT-3 on a dataset minimally processed to standardize pronouns and remove frequent stop words, with an 80%-20% training-testing split. | GPT-3 was tested on three different dialectal Malay datasets (combined, KL, and non-KL), performing best on the KL dataset with a max_example value of 10, which achieved the highest overall performance. Despite the promising results, the non-KL dataset showed the lowest performance, suggesting that larger or more homogeneous datasets might be necessary for improved accuracy in depression detection tasks. |
| (Wang et al., 2020) [62] | Detecting depression using LLMs through microblogs | BERT; RoBERTa [51]; XLNet [63] | 13,993 microblogs collected from the Sina Weibo [64] | The study utilized a dataset from Sina Weibo, annotated for varying levels of depression risk, to evaluate the effectiveness of BERT, RoBERTa, and XLNET models in classifying depression risk through fine-tuning and domain-specific pretraining. | RoBERTa achieved the highest macro-averaged F1 score of 0.424 for depression classification, while BERT scored the highest micro-averaged F1 score of 0.856. Pretraining on an in-domain corpus improved model performance. |

| Reference | Task | Model | Dataset | Methodology | Results |
|---|---|---|---|---|---|
| (Metzler et al., 2022) [65] | Detecting suicidal ideation using LLMs through Twitter texts | BERT; XLNet [63] | 3202 English tweets | The study involved manually labeling 3,202 English tweets according to a novel scheme, leading to training various machine learning models, including deep learning (BERT, XLNet), for multiclass and binary classification tasks aimed at suicide prevention. | BERT achieved F1-scores of 0.93 for accurately labeling tweets as about suicide and 0.74 for off-topic tweets in the binary classification task. Its performance was similar to or exceeded human performance and matched that of state-of-the-art models on similar tasks. |
| (Sadeghi et al., 2023) [66] | Detecting depression using LLMs through interviews | GPT-3.5-Turbo; RoBERTa [51] | E-DAIC (219 participants) [67] | The study utilized GPT-3.5-Turbo to transform interview transcripts, making them more informative for detecting depression. It then employed the DepRoBERTa language model, fine-tuned on these transformed transcripts, to predict an individual's Patient Health Questionnaire (PHQ) score based on text analysis. | The study achieved its lowest error rates, a Mean Absolute Error (MAE) of 3.65 on the dev set and 4.26 on the test set, by fine-tuning DepRoBERTa with a specific prompt, outperforming manual methods and highlighting the potential of automated text analysis for depression detection. |
| (Zhang et al., 2021) [68] | Detecting depression trends using LLMs through Twitter texts | RoBERTa [51]; XLNet [63] | 2575 Twitter users with depression identified via tweets and profiles | The study identified depression-related content on Twitter using regular expressions, built a dataset of 2575 users, trained transformer-based models to classify depression, explored a fusion classifier, and demonstrated the model's ability to monitor depression trends during COVID-19. | This study developed a fusion model that accurately classified depression among Twitter users with 78.9% accuracy. It identified key linguistic and behavioral indicators of depression and demonstrated that depressive users responded to the pandemic later than controls. The findings suggest the model's effectiveness in noninvasively monitoring mental health trends during major events like COVID-19. |
| (Vajre et al., 2021) [69] | Detecting mental health using LLMs through social media texts | PsychBERT | Twitter hashtags and Subreddit (6 domains: anxiety, mental health, suicide, etc) | The paper developed a taxonomy based on HiTOP, implemented a two-stage framework for mental health text identification and behavior detection, and incorporated interpretability components. | The study identified PsychBERT as the highest-performing model, achieving an F1 score of 0.98 in a binary classification task and 0.63 in a more challenging multi-class classification task, indicating its superiority in handling complex mental health-related data. Additionally, PsychBERT's explainability was enhanced by using the Captum library, which confirmed its ability to accurately identify key phrases indicative of mental health issues. |
| (Levkovich & Elyoseph, 2023) [70] | Detecting suicidal ideation using LLMs through text vignette | ChatGPT-3.5; ChatGPT-4 | ChatGPT's response to the text vignette from Levi-Belz and Gamliel [71] | ChatGPT-4 and ChatGPT-3.5 evaluated a vignette depicting suicide risk, compared to mental health professionals' assessments. | ChatGPT-4's assessments of suicide attempts aligned closely with mental health professionals with an average Z score of 0.01, while ChatGPT-3.5 significantly underestimated these risks with a Z score of -0.83. ChatGPT-4 reported higher rates of suicidal ideation and psychache with Z scores of 0.47 and 1.00, respectively, but assessed resilience levels lower than professionals with Z scores of -0.89 and -0.90. |
| (Howard et al., 2020) [72] | Detecting suicidal ideation using LLMs through social media texts | DeepMoji [73]; Universal Sentence Encoder [74]; GPT-1 | 1588 labeled posts from the Computational Linguistics and Clinical Psychology 2017 shared task | The study utilized sentiment analysis and linguistic tools, along with pre-trained neural network models, to process 1588 posts from a clinical psychology forum. It then used automated machine learning to generate classifiers for efficiently categorizing these posts. | The top-performing system, utilizing features derived from the GPT-1 model fine-tuned on over 150,000 unlabeled Reachout.com posts, achieved a new state-of-the-art macro-averaged F1 score of 0.572 on the CLPsych 2017 task without relying on metadata or preceding posts. However, error analysis indicated that this system frequently misses expressions of hopelessness. |
| (Stigall et al., 2024) [75] | Emotion Classification using LLMs | EmoBERTTiny | A collection of publicly available | This paper used a parallel multi-task learning approach with a single loss function for Emotion | EmoBERTTiny outperformed pre-trained and state-of-the-art models in all metrics and computational |

| | through social media texts | | datasets hosted on Kaggle and Huggingface [76, 77] | Classification and Sentiment Analysis. It analyzed the fine-tuned BERTTiny model, EmoBERTTiny, comparing its performance to baseline models and 7B parameter models, benchmarking it against Llama-2-7B-chat and Mistral-7B-Instruct in terms of accuracy, F1-score, precision-recall curves, and inference speed. | efficiency, achieving 93.14% accuracy in sentiment analysis and 85.48% in emotion classification. It processes a 256-token context window in 8.04ms post-tokenization and 154.23ms total processing speed. |
|---|---|---|---|---|---|
| (Ghanadian et al., 2024) [78] | Suicidal ideation detection using LLMs through social media texts | ALBERT; DistilBERT; ChatGPT; Flan-T5 [79]; Llama | UMD Dataset [80]; Synthetic Datasets (Generated using LLMs like Flan-T5 and Llama2, these datasets augment the UMD dataset to enhance model performance) | This paper detailed a methodology that first extracted social factors from psychology literature to inform GLLM-based data synthesis prompts. It then used three GLLMs to generate synthetic data on suicide-related topics and trained classifiers on real, synthetic, and augmented datasets, testing their performance on both real and synthetic test sets. | The synthetic data-driven method achieved consistent F1-scores of 0.82, comparable to real-world data models yielding F1-scores between 0.75 and 0.87. When 30% of the real-world UMD dataset was combined with the synthetic data, the performance significantly improved, reaching an F1-score of 0.88 on the UMD test set. This result highlights the effectiveness of synthetic data in addressing data scarcity and enhancing model performance. |
| (Lossio-Ventura et al., 2024) [81] | Evaluations of LLMs for sentiment analysis through social media texts | ChatGPT; Open Pre-Trained Transformers (OPT) | NIH Data Set [82]; Stanford Data Set [83] | This paper created gold standard labels for a subset of each dataset using a panel of human raters. It compared 8 state-of-the-art sentiment analysis tools on both datasets to evaluate variability and disagreement. Additionally, it explored few-shot learning by fine-tuning OPT using a small annotated subset and zero-shot learning using ChatGPT. | This paper revealed high variability and disagreement among sentiment analysis tools when applied to health-related survey data. OPT and ChatGPT demonstrated superior performance, outperforming all other tools. Moreover, ChatGPT outperformed OPT, achieving 6% higher accuracy and a 4% to 7% higher F-measure. |

**Table 2. Summary of the 7 selected articles from the literature on LLMs in mental health CAs.**

| Ref. | Cases | Models | Data Sources | Methodology Used | Main Outcomes |
|---|---|---|---|---|---|
| (Beredo & Ong, 2022) [84] | Mental health interventions using CAs supported by LLMs | EREN [85]; MHBot [86]; PERMA [87] | Empatheticdialogues (24,850 conversations) [88]; Well-Being Conversations [89]; Perma Lexica [90] | This study evaluated LLMs involves automated evaluation, where metrics like perplexity measure a model's ability to predict unseen test sets, and human evaluation, assessing the chatbot's human-likeness and response quality through criteria like performance, humanity, and affect, evaluated by experts in psychology. Additionally, experts were recruited to assess chatbot interactions based on specific quality attributes, providing a comprehensive understanding of the model's conversational abilities. | This study successfully demonstrated a hybrid conversation model, which combines generative and retrieval approaches to improve language fluency and empathetic response generation in chatbots. This model, tested through both automated metrics and human evaluation, showed that the medium variation of the FTER model outperformed the vanilla DialoGPT in perplexity and that the human-likeness, relevance, and empathetic qualities of responses were significantly enhanced, making VHope a more competent CA with empathetic abilities. |
| (Crasto et al., 2021) [91] | Mental health interventions using CAs supported by LLMs | DialoGPT | Counselchat (includes tags of illness); question answers from 100 college students | Recognized mental health questionnaires (PHQ-9 & WHO-5) were completed. The DialoGPT fine-tuned with Counselchat data, was employed for chatbot interaction. Micro-interventions were suggested based on identified issues, and a student survey was administered. | The DialoGPT model, demonstrating higher perplexity and preferred by 63% of college participants for its human-like and empathetic responses, was chosen as the most suitable system for addressing student mental health issues. |

| Reference | Topic | Tools/Models | Data | Methodology | Findings |
|---|---|---|---|---|---|
| (Zygadlo, 2021) [92] | Mental health interventions using Polish-language CA supported by LLMs | Rasa [93]; spaCy [94]; Transformers; BERT | EmpatheticDialogues [88]; DailyDialog [95] | The paper entailed developing a chatbot with Rasa and an emotion recognition model, creating a bilingual Polish-English corpus from existing datasets, and employing machine translation for Polish. This approach facilitated sentiment and emotion classification using BERT models, demonstrating the effective use of machine translation for data-scarce languages. | The successful setup of an initial chatbot dialogue framework using Rasa and the development of a bilingual (English and Polish) corpus for emotion recognition. The research has advanced to training BERT-based models for emotion recognition, achieving high accuracy in sentiment and emotion classification, demonstrating the feasibility of integrating machine translation to work with less-resourced languages like Polish for emotional understanding in chatbots. |
| (Ma et al., 2024) [14] | Evaluation of mental health intervention CAs supported by LLMs | GPT-3 | 120 Reddit posts (2913 user comments) | The study utilized a qualitative content analysis of Reddit posts from the r/Replika subreddit to explore user experiences with the AI-based CA Replika, focusing on mental well-being support. By employing a two-stage coding process with a developed codebook, researchers analyzed a representative sample of posts and comments to identify key benefits and challenges. | The study highlighted that CAs like Replika, powered by LLMs, offered crucial mental health support by providing immediate, unbiased assistance and fostering self-discovery. However, they struggled with content filtering, consistency, user dependency, and social stigma, underscoring the importance of cautious use and improvement in mental wellness applications. |
| (Heston, 2023) [96] | Evaluation of mental health intervention CAs supported by LLMs | ChatGPT-3.5 | Public AI mental health CAs from FlowGPT.com | Evaluated ChatGPT-3.5 mental health agents with simulations for recognizing suicidality, tracking referral to humans, and shutdown at risk levels. | This study evaluated 25 Cas from FlowGPT.com, finding that they referred to human intervention at moderate depression levels (PHQ-9 score of 12) and shut down at severe levels (score of 25). Only two agents provided crisis resources, and most resumed dialogue if the risk level decreased. |
| (Alessa and Al-Khalifa, 2023) [97] | Mental health interventions using CAs for the elderly supported by LLMs | ChatGPT; Google Cloud API | Record of interactions with CA; results of the human experts' assessment | This paper explored using ChatGPT to create a chatbot for providing support to older adults and socially isolated seniors. The system incorporated Google's Cloud API for speech recognition and text-to-speech, and personalized prompts based on user information collected through a questionnaire. The chatbot engaged users in empathetic conversations, quizzes, and health tips, with prompt engineering optimized through three experiments. | The proposed ChatGPT-based system effectively serves as a companion for elderly individuals, helping to alleviate loneliness and social isolation. Preliminary evaluations showed that the system could generate relevant responses tailored to elderly personas. |
| (He et al., 2024) [98] | Evaluation of CAs handling counseling for people with autism supported by LLMs | ChatGPT | Public available data from the web-based medical consultation platform DXY [99] | This paper selected 100 patient consultation samples related to autism from January 2018 to August 2023. The questions and responses were anonymized and randomized. Three chief physicians assessed the responses across four dimensions: relevance, accuracy, usefulness, and empathy, completing 717 evaluations. The responses were then compared using a Likert scale to gauge their quality. | The study found that 46.86% of assessors preferred responses from physicians, 34.87% favored ChatGPT, and 18.27% favored ERNIE Bot. Physicians and ChatGPT showed higher accuracy and usefulness compared to ERNIE Bot, while ChatGPT outperformed both in empathy. The study concluded that while physicians' responses were generally superior, LLMs like ChatGPT can provide valuable guidance and greater empathy, though further optimization and research are needed for clinical integration. |

**Table 3. Summary of the 18 selected articles from the literature on other applications and evaluation of the LLMs in mental health.**

| Ref. | Cases | Model | Data Sources | Methodology Used | Main Outcomes |
|---|---|---|---|---|---|
| (Franco D'Souza et al., 2023) [100] | Evaluation of ChatGPT's responses to clinical vignettes in psychiatry | ChatGPT 3.5 | 100 Cases in Psychiatry [101]; ChatGPT 3.5 responses to cases | ChatGPT 3.5 responded to 100 psychiatric case vignettes, evaluated by expert faculties across 10 categories using mean scores, and represented graphically. | ChatGPT 3.5 received mostly "Grade A" ratings in 61 out of 100 cases, excelling in management strategies and diagnoses across psychiatric conditions. Few responses received "Grade C" due to minor discrepancies, but no diagnostic errors were noted. |
| (Spallek et al., 2023) [102] | Evaluation of ChatGPT in mental health education | GPT-4 | Real-world data from 'Cracks in the Ice' [103] and 'Positive Choices' [104] | The study utilized GPT-4 and real-world queries and factsheets from two health portals, assessing LLMs' potential in generating mental health education content within ethical guidelines. | GPT-4's outputs seemed valid but were substandard compared to expert materials, lacking in reading level and adherence to guidelines, requiring careful human editing. Although not suitable for direct consumer queries, GPT-4 can be cautiously used by educators and researchers to develop educational materials, which should disclose AI use and be evaluated for efficacy. |
| (Farhat et al., 2023) [105] | Evaluation of ChatGPT as a complementary mental health resource | ChatGPT | Responses generated by ChatGPT | The study evaluated ChatGPT's effectiveness in mental health support by analyzing its responses and cross-questioning, particularly focusing on issues related to anxiety and depression and its suggestions regarding medications. | ChatGPT displayed significant inconsistencies and low reliability when providing mental health support for anxiety and depression, underlining the necessity of validation by medical professionals and cautious use in mental health contexts. |
| (Wei et al., 2023) [106] | Evaluation of ChatGPT in psychiatry | ChatGPT | Theoretical analysis and literature reviews | The study investigated ChatGPT's application in psychiatry, evaluating its capabilities in screening, diagnosis, and patient support. | The paper found ChatGPT useful in psychiatry, stressing ethical use and human oversight, while noting challenges in accuracy and bias, positioning AI as a supportive tool in care. |
| (Yongsatianchot et al., 2023) [107] | Evaluation of LLMs' perception of emotion | Text-davinci-003 [108]; ChatGPT; GPT-4 | Responses from three OpenAI LLMs to the Stress and Coping Process Questionnaire | The study assessed the emotional understanding of LLMs like ChatGPT using the Stress and Coping Process Questionnaire (SCPQ) across three OpenAI models (davinci-003, ChatGPT, GPT-4) to compare their appraisal and coping reactions against human data and appraisal theory predictions. | The study applied the SCPQ to three OpenAI LLMs—davinci-003, ChatGPT, and GPT-4—and found that while their responses aligned with human dynamics of appraisal and coping, they did not vary across key appraisal dimensions as predicted and differed significantly in response magnitude. Notably, all models reacted more negatively than humans to negative scenarios, potentially influenced by their training processes. |
| (Grabb, 2023) [109] | Evaluation of prompt engineering by LLMs and its impact on mental health | ChatGPT 4.0 | ChatGPT's answers to 4 unique questions | The study tested ChatGPT 4.0's response variability to four uniquely framed questions about happiness, each asked five times in distinct roles and contexts, to explore the model's adaptability and advice consistency. | The study found ChatGPT 4.0's advice varied widely based on prompt design, emphasizing careful prompt crafting in mental healthcare contexts to ensure safety and relevance. |
| (Hadar-Shoval et al., 2023) [110] | Evaluation of ChatGPT's mentalizing abilities in borderline personality disorder (BPD) and schizoid | ChatGPT 3.5 | Rating of Levels of Emotional Awareness Scale (LEAS) scenarios for BPD and SPD by ChatGPT | The study evaluated ChatGPT's emotional awareness through modified LEAS scenarios for BPD and SPD, scoring responses and analyzing differences in emotion identification and intensity. | ChatGPT was able to accurately describe the emotional reactions of individuals with BPD as more intense, complex, and rich than those with SPD. |
| | personality disorder (SPD) | | | | |
| (Sezgin et al., 2023) [111] | Evaluation of clinical accuracy in LLMs' responses to postpartum depression (PPD) questions | GPT-4 (using ChatGPT); LaMDA (using Bard) [112] | 14 PPD-related patient-focused frequently asked questions sourced from the American College of Obstetricians and Gynecologists | The study compared responses from GPT-4, LaMDA, and Google Search against ACOG's FAQs on postpartum depression, evaluated by two board-certified physicians using a GRADE-informed scale. Statistical analyses were performed using R software, including interrater reliability and differences in response quality. | ChatGPT outperformed Bard and Google Search in providing high-quality, clinically accurate responses to postpartum depression questions, with significant statistical support and perfect rater agreement on its responses. |
| (Tanana et al., 2021) [113] | Evaluation of LLM's ability to rate emotions in psychotherapy | BERT; LIWC [114] | Psychotherapy transcripts that were published by Alexander Street Press [115]; the human ratings from a database of 97,497 utterances from psychotherapy | The paper utilized psychotherapy transcripts to extract utterances for sentiment analysis, employing N-gram models, a recurrent neural network, LIWC, and BERT for comparison. Evaluation metrics included overall accuracy, F1 score, and Cohen's kappa. | MaxEnt models surpassed LIWC, with BERT achieving the highest performance (kappa = 0.48). The best model exceeded human performance on the test set by 14%. |
| (Wang et al., 2020) [116] | Enhancing depression diagnosis and treatment through the use of LLMs | LLaMA-7B; ChatGLM-6B; Alpaca; LLMs+Knowledge | Chinese Incremental Pre-training dataset [117] | The paper customized a Chinese language model for depression, using datasets and a knowledge graph. Techniques included data augmentation, generating instruction data from the knowledge graph, fine-tuning the model, and reinforcement learning with expert feedback. | The study assessed LLMs' performance in mental health, emphasizing safety, usability, and fluency and integrating mental health knowledge to improve model effectiveness, enabling more tailored dialogues for treatment while ensuring safety and usability. |
| (Schubert et al., 2023) [118] | Evaluation of LLMs' performance on neurology board-style examinations | ChatGPT 3.5; ChatGPT 4.0 | A question bank from an educational company with 2036 questions that resemble neurology board questions [119] | The study evaluated two LLMs using a neurology question bank, categorizing questions into lower and higher-order types. Statistical analysis compared model performance with human performance. | ChatGPT 4.0 excelled over ChatGPT 3.5, achieving 85.0% accuracy versus 66.8%. It surpassed human performance in specific areas and exhibited high confidence in responses. Longer questions tended to result in more incorrect answers for both models. |
| (Friedman & Ballentine, 2023) [120] | Evaluation of LLMs in data-driven discovery: correlating sentiment changes with psychoactive experiences | BERTowid [29]; BERTiment [121] | Erowid testimonials [122]; drug receptor affinities [123]; brain gene expression data [124]; 58K annotated Reddit posts [125] | This paper used BERT and 11,816 testimonials to predict sentiments and demographics, then linked drug effects to words, identifying 11 key factors on a 3D brain model. | This paper found that LLM methods can create unified and robust quantifications of subjective experiences across various psychoactive substances and timescales. The representations learned are evocative and mutually confirmatory, indicating significant potential for LLMs in characterizing psychoactivity. |
| (Wu et al., 2023) [126] | Expanding dataset of Post-Traumatic Stress Disorder (PTSD) using LLMs | GPT- 3.5 Turbo | E-DAIC (219 participants) [67] | This paper developed two text augmentation frameworks utilizing LLMs to address data imbalance in NLP tasks for PTSD diagnosis. The methodologies applied were zero-shot, which generated standardized transcripts, and few-shot, which rephrased existing training samples within the E-DAIC. | This paper demonstrated that two novel text augmentation frameworks using LLMs significantly improved PTSD diagnosis by addressing data imbalances in NLP tasks. The zero-shot approach, which generated new standardized transcripts, achieved the highest performance improvements, while the few-shot approach, which rephrased existing training samples, also surpassed the original dataset's efficacy. |
| (Kumar et al., 2023) 127 | Evaluation of GPT 3 in mental health intervention | GPT 3 | 209 participants responses, with 189 valid responses after filtering | This paper conducted a pilot experiment using a 2x2x2 factorial design to compare LLM-based chatbot interventions with video-based methods for improving mental health awareness. GPT-3 was used to create chatbots for providing mindfulness information and reflection, and participants were | This paper found that interaction with either of the chatbots improved participants' intent to practice mindfulness again, while the tutorial video enhanced their overall experience of the exercise. These findings highlighted the potential promise and outlined directions for exploring the use of LLM-based |

| | | | | recruited from Amazon Mechanical Turk. | | chatbots for awareness-related interventions. |
|---|---|---|---|---|---|---|
| (Elyoseph et al., 2024) [128] | Evaluation of LLMs in mental health intervention | ChatGPT3.5; ChatGPT4; Claude; Bard | ChatGPT 3.5, ChatGPT 4, Claude, Bard, and mental health professionals' responses to text vignettes about depression | This paper conducted a comparative analysis using case vignettes to evaluate the performance of different LLMs against mental health professionals and the general public. The focus was on the LLMs' ability to generate prognoses, anticipated outcomes with and without intervention, and long-term consequences for individuals with depression. | | This paper found that ChatGPT-4, Claude, and Bard aligned closely with mental health professionals and the general public in diagnosing depression and recommending combined treatment, while ChatGPT-3.5 had a more pessimistic prognosis. The study highlighted AI's potential to complement mental health professionals but raised concerns about ChatGPT-3.5's impact on patient motivation. |
| (Perlis et al., 2024) [129] | Evaluation of GPT-4 for clinical decision support in bipolar depression | GPT-4 turbo (gpt-4-1106-preview) | Recommendations generated by the augmented GPT-4 model and responses from clinicians treating bipolar disorder | This paper generated 50 vignettes of bipolar disorder cases and had expert clinicians rank treatment options. It then compared these rankings with recommendations from an augmented GPT-4 model using specific guidelines and also evaluated responses from a community clinician group. | | This paper found that the augmented GPT-4 model had a Cohen's kappa of 0.31 with expert consensus, identifying the optimal treatment in 50.8% of cases and placing it in the top 3 in 84.4% of cases. In contrast, the base model had a Cohen's kappa of 0.09 and identified the optimal treatment in 23.4% of cases, highlighting the enhanced performance of the augmented model in aligning with expert recommendations. |
| (Blease et al., 2024) [130] | Evaluation of psychiatrists' perceptions of the LLMs | ChatGPT; Bard; Bing AI | Survey responses from 138 APA members on LLM chatbot use in psychiatry | This paper surveyed APA members who attended an "AI in Psychiatry" informational session to explore their experiences and opinions on using LLM-powered chatbots in clinical practice. Participants provided feedback through a three-minute survey divided into sections on chatbot usage, its effects on clinical practice, and patient interactions, with responses analyzed using descriptive statistics and thematic analysis. | | This paper found that over half of psychiatrists used AI tools like ChatGPT for clinical questions, with nearly 70% agreeing on improved documentation efficiency and almost 90% indicating a need for more training while expressing mixed opinions on patient care impacts and privacy concerns. |
| (Berrezueta-Guzman et al., 2024) [131] | Evaluation of the efficacy of ChatGPT in mental intensive treatment | ChatGPT | Evaluations from 10 attention deficit hyperactivity disorder (ADHD) therapy experts and interactions between therapists and the custom ChatGPT | This paper developed a custom ChatGPT based on a literature review and validated it with therapeutic experts before implementing it in a robotic assistant for ADHD therapies. The Delphi method was used with a panel of ten experts to assess the ChatGPT's performance across various therapeutic categories, ensuring a thorough, expert-driven evaluation. | | This paper found that the custom ChatGPT demonstrated strong capabilities in engaging language use, maintaining interest, promoting active participation, and fostering a positive atmosphere in ADHD therapy sessions, with high ratings in communication and language. However, areas needing improvement were identified, particularly in confidentiality and privacy, cultural and sensory sensitivity, and handling nonverbal cues. |

## 3.2 Mental health conditions and suicidal ideation detection through text

Early intervention and screening are crucial in mitigating the global burden of mental health issues [132]. We examined the performance of LLMs in detecting mental health conditions and suicidal ideation through textual analysis. Six articles assessed the efficacy of early screening for depression using LLMs [50,57,60,61,66,68], while another simultaneously addressed both depression and anxiety [60]. One comprehensive study examined various psychiatric conditions, including depression, social anxiety, loneliness, anxiety, and other prevalent mental health issues [69]. Two articles assessed and compared the ability of LLMs to perform sentiment and emotion analysis [75,81]. Five articles focused on the capability of LLMs to analyze textual content for detecting suicidal ideation [54,65,70,72,78]. Most studies employed BERT and its variants as one of the primary models (n=10) [50,54,57,62,65,66,68,69,75,78], while GPT models were also commonly used (n=8) [57,60,61,66,70,72,78,81]. The majority of training data comprised social media posts (n=10) [50,54,62,65,68,69,72,75,78,81] from platforms like Twitter, Reddit, and Sina Weibo, covering languages such as English, Malay dialects, Chinese, and Portuguese. Additionally, five studies utilized datasets consisting of clinical transcripts and patient interviews [50,57,60,61,66], providing deeper insights into LLM applications in clinical mental health settings.

In studies focusing on early screening for depression, comparing results horizontally is challenging due to variations in datasets, training methods, and models across different investigations. Nonetheless, substantial evidence supports the significant potential of LLMs in detecting depression from text-based data. For example, Danner et al. conducted a comparative analysis using a Convolutional Neural Network (CNN) on the DAIC-WOZ dataset, achieving F1 scores of 0.53 and 0.59; however, their use of GPT-3.5 demonstrated superior performance, with an F1 score of 0.78 [57]. Another study involving the E-DAIC dataset (an extension of DAIC-WOZ) used DepRoBERTa to predict the PHQ-8 scores from textual data. This approach identified three levels of depression and achieved the lowest MAE of 3.65 in PHQ-8 scores [66].

LLMs play an important role in sentiment analysis [75,81], which categorizes text into overall polarity classes such as positive, neutral, negative, and occasionally mixed, and emotion classification, which assigns labels like 'joy,' 'sadness,' 'anger,' and 'fear' [75]. These analyses enable the detection of emotional states and potential mental health issues from textual data, facilitating early intervention [133]. Stigall et al. demonstrated the efficacy of these models, with their study showing that EmoBERTTiny, a fine-tuned variant of BERT, achieved an accuracy of 93.14% in sentiment analysis and 85.46% in emotion analysis. This performance surpasses that of baseline models, including BERT-Base Cased and Prak-wal1 pre-trained BERTTiny [75], underscoring the advantages and validity of fine-tuning in enhancing model performance. LLMs have also demonstrated robust accuracy in detecting and classifying a range of mental health syndromes, including social anxiety, loneliness, and generalized anxiety. Vajre et al. introduced

PsychBERT, developed using a diverse training dataset from both social media texts and academic literature, which achieved an F1 score of 0.63, outperforming traditional deep learning approaches such as CNNs and Long Short-Term Memory Networks (LSTMs), which recorded F1 scores of 0.57 and 0.51, respectively [69]. In research on detecting suicidal ideation using LLMs, Diniz et al. showcased the high efficacy of the BERTimbau Large model within a non-English (Portuguese) context, achieving an accuracy of 0.955, precision of 0.961, and an F-score of 0.954 [54]. Metzler et al.'s assessment of the BERT model found it correctly identified 88.5% of tweets as suicidal or off-topic, performing comparably to human analysts and other leading models [65]. However, Inbar Levkovich et al. noted that while ChatGPT-4 assessments of suicide risk closely aligned with those by mental health professionals, it overestimated suicidal ideation [70]. These results underscore that while LLMs have the potential to identify tweets reflecting suicidal ideation with accuracy comparable to psychological professionals, extensive follow-up studies are required to establish their practical application in clinical settings.

### 3.3 LLMs in mental health CAs

In the growing field of mental health digital support, the implementation of LLMs as CAs has exhibited both promising advantages [14,84,91,96] and significant challenges [92,96]. The studies by Ma et al. and Heston et al. both demonstrate the effectiveness of CAs powered by LLMs in providing timely, non-judgmental mental health support [14,96]. This intervention is particularly important for those who lack ready access to a therapist due to constraints such as time, distance, and work, as well as for certain socially marginalized populations, such as older adults who experience chronic loneliness and a lack of companionship [14,97]. Ma et al.'s qualitative analysis of user interactions on Reddit highlights that LLMs encourage users to speak up and boost their confidence by providing personalized and responsive interactions [14]. Additionally, VHope, a DialoGPT-enabled mental health CA, was evaluated by three experts who rated its responses as 67% relevant, 78% human-like, and 79% empathetic [84]. Another study found that after 717 evaluations by 100 participants on 239 autism-specific questions, 46.86% of evaluators preferred responses of the chief physicians, whereas 34.87% preferred ChatGPT-4 (OpenAI), and 18.27% favored ERNIE Bot (version 2.2.3; Baidu, Inc). Moreover, ChatGPT (mean score: 3.64, 95% CI 3.57-3.71) outperformed physicians (mean score: 3.13, 95% CI 3.04-3.21) in terms of empathy [98], indicating that LLM-powered CAs are not only effective but also acceptable by users. These findings highlight the potential for LLMs to complement mental health intervention systems and provide valuable medical guidance.

The development and implementation of a non-English CA for emotion capture and categorization was explored in a study by Zygadlo et al. Faced with a scarcity of Polish datasets, the study adapted by translating an existing database of personal conversations from English into Polish, which decreased accuracy in tasks from 90% in English to 80% in Polish [92]. While the performance remained commendable, it highlighted the challenges posed by the lack of robust datasets in languages other than English, impacting the effectiveness of CAs across different linguistic environments. However, findings by He et al. suggest that the availability of language-specific datasets is not the sole determinant of CA performance. In their study, although ERNIE Bot was trained in Chinese and ChatGPT in English, ChatGPT demonstrated greater empathy for Chinese users [98]. This implies that factors beyond the training language and dataset availability, such as model architecture or training methodology, can also affect the empathetic responsiveness of LLMs, underscoring the complexity of human-AI interaction.

Meanwhile, the reliability of LLM-driven CAs in high-risk scenarios remains a concern [14,96]. An evaluation of 25 CAs found that in tests involving suicide scenarios, only two included suicide hotline referrals during the conversation [96]. This suggests that while these CAs can detect extreme emotions, few are equipped to take effective preventive measures. Furthermore, CAs often struggle with maintaining consistent communication due to limited memory capacity, leading to disruptions in conversation flow and negatively affecting user experience [14].

### 3.4 The other applications and evaluation of the LLMs in mental health

ChatGPT has gained attention for its unparalleled ability to generate human-like text and analyze large amounts of textual data, attracting the interest of many researchers and practitioners [100]. Numerous evaluations of LLMs in mental health have focused on ChatGPT, exploring its utility across various scenarios such as clinical diagnosis [100,106,111], treatment planning [106,128,131], medication guidance [105,109,129], patient management [106], psychiatry examinations [118], and psychology education [102], among others [107,110,127,130].

Research has highlighted ChatGPT's accuracy in diagnosing various psychiatric conditions [106,110,111,126]. For example, Franco D'Souza et al. evaluated ChatGPT's responses to 100 clinical psychiatric cases, awarding it an 'A' rating in 61 cases, with no errors in the diagnoses of different psychiatric disorders and no unacceptable responses, underscoring ChatGPT's expertise and interpretative capacity in psychiatry [100]. Further supporting this, research from Schubert et al. assessed ChatGPT 4.0's performance using neurology board-style exam questions, finding that it answered 85% of the questions correctly, surpassing the average human performance of 73.8% [118]. Meanwhile, in a study of LLMs regarding the prognosis and long-term outcomes of depression, ChatGPT-4, Claude, and Bard showed strong agreement with mental health professionals. They all recommended a combination of psychotherapy and antidepressant medication in every case [130]. This not only proves the reliability of LLMs for mental health assessment but also highlights their usefulness in providing valuable support and guidance for individuals seeking information or coping with mental illness.

However, the direct deployment of LLMs such as ChatGPT in clinical settings carries inherent risks. The outputs of LLMs are heavily influenced by prompt engineering, which can lead to inconsistencies that undermine clinical reliability [102,105,106,107,109]. For example, Farhat et al. conducted a critical evaluation of ChatGPT's ability to generate medication guidelines through detailed cross-questioning and noted that altering prompts substantially changed the responses [105]. While ChatGPT typically provided helpful advice and recommended seeking expert consultation, it occasionally produced inappropriate medication suggestions. Perlis et al. verified this, showing that GPT-4 Turbo suggested medications that were considered poor choices or contraindicated by experts in 12% of cases [129]. Moreover, LLMs often lack the necessary clinical judgment capabilities. This issue was highlighted by Grabb's study, which revealed that despite built-in safeguards, ChatGPT remains susceptible to generating extreme and potentially hazardous recommendations [109]. A particularly alarming example was ChatGPT advising a depressed patient to engage in high-risk activities like bungee jumping as a means of seeking pleasure [109]. These LLMs depend on prompt engineering [102,105,109], which means their responses can vary widely depending on the wording and context of the prompts given. The system prompts, which are predefined instructions given to the model, and the prompts used by the experimental team, such as those in Farhat's study, guide the behavior of ChatGPT and similar LLMs. These prompts are designed to accommodate a variety of user requests within legal and ethical boundaries. However, while these boundaries are intended to ensure safe and appropriate responses, they often fail to align with the nuanced sensitivities required in psychological contexts. This mismatch underscores a significant deficiency in the clinical judgment and control of LLMs within sensitive mental health settings.

Further research into other LLMs in the mental health sector has shown a range of capabilities and limitations. For example, a study by Sezgin et al. highlighted LaMDA's proficiency in managing complex inquiries about PPD that require medical insight or nuanced understanding, yet pointed out its challenges with straightforward, factual questions, such as "What are antidepressants?" [111]. Assessments of LLMs like LLaMA-7B, ChatGLM-6B, and Alpaca, involving 50 interns specializing in mental illness, received favorable feedback regarding the fluency of these models in a clinical context, with scores above 9.5 out of 10. However, the results also indicated that the responses of these LLMs often failed to address mental health issues adequately, demonstrated limited professionalism, and resulted in decreased usability [116]. Similarly, a study on psychiatrists' perceptions of using LLMs such as Bard and Bing AI in mental health care revealed mixed feelings. While 40% of physicians indicated that they would use such LLMs to assist in answering clinical questions, some expressed serious concerns about their reliability, confidentiality, and potential to damage the patient-physician relationship [130].

**Table 4. Summary of main strengths, limitations, and suggestions of LLMs in mental health from the selected articles.**

| CATEGORY | STRENGTH | LIMITATION | SUGGESTION |
|---|---|---|---|
| **Mental Illness and Suicidal Ideation Detection** | • High Detection Accuracy: Advanced LLMs achieve high accuracy in detecting depression and other mental health issues from textual data [50,57].<br>• Early Detection Capabilities: LLMs effectively enable early detection of suicidal ideation and depression, crucial for timely interventions [54,66,78].<br>• Multilingual and Diverse Data Handling: Demonstrated capability of LLMs in handling data across different languages and cultural contexts, improving global mental health monitoring [60,61]. | • Privacy and Ethical Concerns: Ethical concerns due to the passive collection of sensitive text data, underscore the need for improved privacy controls [54,60].<br>• Contextual and Emotional Understanding Deficits: Limitations in missing deeper contextual and emotional nuances critical for accurate mental health assessments [60,65,75].<br>• Generalizability Challenges: Challenges in generalizing across diverse populations due to training on specific datasets [65,67,78,81]. | • Improve Privacy and Ethics: Focus on enhancing privacy controls and adhering to ethical standards in future developments [54,60].<br>• Enhance Contextual Understanding: Incorporate multimodal data to improve models' understanding of context and emotions [60,65].<br>• Expand Cross-cultural and Multilingual Research: Advance models with diverse datasets to improve universality and adaptability [60,61,78]. |
| **Mental Health CAs** | • Enhanced Empathy and Engagement: LLMs like VHope and Replika offer empathetic and human-like interactions [14,84,98].<br>• Accessibility and Reach: LLMs can provide support at scale and across different languages, making mental health support more accessible to diverse populations [92,97].<br>• Complex Inquiry Handling: LLMs are adept at handling complex medical and emotional inquiries, providing nuanced responses based on extensive datasets [84,98]. | • Inconsistency and Reliability: LLMs like ChatGPT can produce inconsistent and sometimes unreliable outputs, especially in high-stakes scenarios such as diagnosing or managing mental health conditions [96].<br>• Lack of Deep Understanding: LLMs struggle with understanding context deeply and can give inappropriate responses, lacking the sensitivity required for certain mental health interactions [14].<br>• Inadequate Crisis Response and Safety Protocols: LLMs may not adequately identify or respond appropriately to severe mental health crises, which can be dangerous if the system fails to refer users to human intervention in a timely and effective manner [96]. | • Enhanced Model Testing and Validation: Implement comprehensive testing protocols and simulations of mental health scenarios to improve the consistency and reliability of LLM outputs [96,97].<br>• Advanced Contextual Understanding: Develop and integrate advanced algorithms for LLMs to enhance their ability to comprehend and respond appropriately to the complex nuances of mental health conversations [92,98].<br>• Robust Crisis Management Integration: Integrate sophisticated algorithms for crisis detection and escalation within LLMs, working in partnership with healthcare professionals to ensure these systems align with clinical standards [96]. |
| **Other Applications of LLMs in Mental Health** | • Diverse Applications: LLMs have shown broad applications in mental health, excelling in diagnostic aid, therapeutic strategy development, and educational material creation, which could enhance both patient care and medical training. [100,102,106,111,118,127,128,129,130,131].<br>• Emotional and Behavioral Insights: Some LLMs effectively mimic and understand human emotional and behavioral patterns, supporting their use in psychotherapy to analyze and respond to patient emotions accurately [107,113,128].<br>• Enhance the Quality and Diversity of Datasets: LLMs can generate synthetic clinical data through advanced text enhancement techniques and knowledge-enhanced training, reducing costs and increasing data availability [116,126]. | • Generalizability and Bias Concerns: The specific dataset or context used and inherent biases in the training data may affect the fairness and accuracy of LLM outputs [109,110,116,127,128,131].<br>• Consistency and Reliability Issues: The outputs of LLMs might vary when generating diagnostic recommendations or treatment strategies, requiring thorough validation and supervision [105,109].<br>• Ethical and Safety Risks: The deployment of LLMs in mental health contexts introduces significant concerns, including the risk of violating data privacy, the potential for generating harmful advice, and the challenges of adhering to ethical standards [102,106,131]. | • Expanding Research Scope and Diversity: Increasing application research in different datasets, different populations, and clinical settings to ensure the generalizability and applicability of LLMs [102,107,127,128,129,131].<br>• Human Oversight and Annotation: Creating specialized, clinically relevant training datasets with expert input and oversight to ensure accurate and stable model outputs [102,105,106,109,129].<br>• Developing Integration and Regulatory Frameworks: Developing detailed guidelines and frameworks to ensure the ethical and safe use of AI in clinical practices and to complement rather than replace human healthcare providers [106,111,131]. |

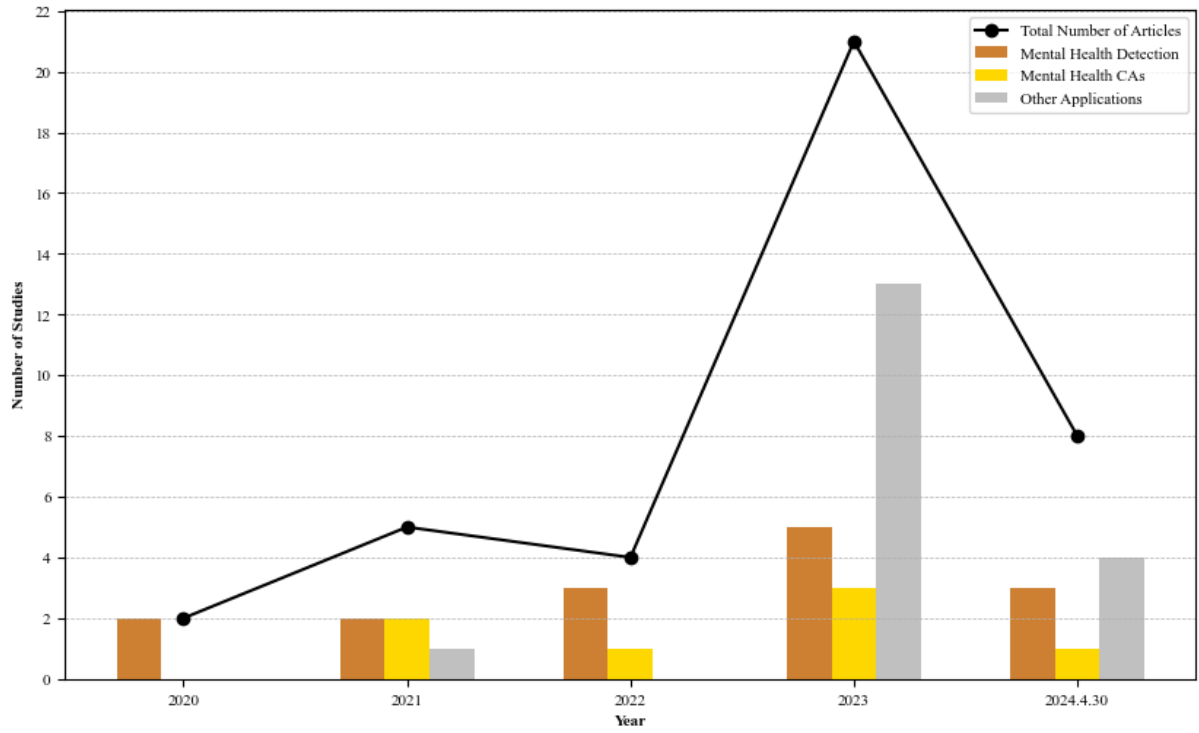

Figure 3. Number of articles included in this literature review, grouped by year of publication and application field. The black line indicates the total number of articles in each year.

## 4. Discussion

### 4.1 Principal findings

In the context of the wider prominence of LLMs in the literature [14,50,57,60,61,69,96,130], our research supports the assertion that interest in LLMs is growing in the field of mental health. Figure 3 indicates a rising trend in the number of mental health studies employing LLMs, with a notable surge observed in 2023 following the introduction of ChatGPT in late 2022. Although we included articles only up to the end of April 2024, it is evident that the number of articles related to LLMs in the field of mental health continues to show a steady increase in 2024. This trend marks a substantial shift in the discourse around LLMs, reflecting their broader acceptance and integration into various aspects of mental health research and practice. The progression from text analysis to a diverse range of applications highlights the academic community's recognition of the multifaceted uses of LLMs. LLMs are increasingly employed for complex psychological assessments, including early screening, diagnosis, and therapeutic interventions.

Our findings demonstrate that LLMs are highly effective in analyzing textual data to assess mental states and identify suicidal ideation [50,54,57,60,61,65,66,68,69,72,78], although their categorization often tends to be binary [50,54,65,68,69,72,78]. These LLMs possess extensive knowledge in the field of mental health and are capable of generating empathic responses that closely resemble human interactions [97,98,107]. They show great potential for providing mental health interventions with improved prognoses [50,96,110,127,128,131], with the majority being recognized by psychologists for their appropriateness and accuracy [98,100,129]. The careful and rational application of LLMs can enhance mental health care efficiently and at a lower cost, which is crucial in areas with limited healthcare capacity. However, there are currently no studies available that provide evaluative evidence to support the clinical use of LLMs.

### 4.2 Limitations

#### 4.2.1 Strengths and Limitations of Using LLMs in Mental Health

Based on the works of literature the strengths and weaknesses of applying the LLMs in mental health are summarized in Table 4.

LLMs have a broad range of applications in the mental health field. These models excel in user interaction, provide empathy and anonymity, and help reduce the stigma associated with mental illness [14,107], potentially encouraging more patients to participate in treatment. They also offer a convenient, personalized, and cost-effective way for individuals to access mental health services at any time and from any location, which can be particularly helpful for socially isolated populations, especially the elderly [60,84,97]. Additionally, LLMs can help reduce the burden of care during times of severe healthcare resource shortages and patient overload, such as during the COVID-19 pandemic [68]. Although previous research has highlighted the potential of LLMs in mental health, it is evident that they are not yet ready for clinical use due to unresolved technical risks and ethical issues.

The use of LLMs in mental health, particularly those fine-tuned for specific tasks such as ChatGPT, reveals clear limitations. The effectiveness of these models heavily depends on the specificity of user-generated prompts. Inappropriate or imprecise prompts can disrupt the conversation's flow and diminish the model's effectiveness [75,96,105,107,109]. Even small changes in the content or tone of prompts can sometimes lead to significant variations in responses, which can be particularly problematic in healthcare settings where interpretability and consistency are critical [14,105,107]. Furthermore, LLMs lack clinical judgment and are not equipped to handle emergencies [95,108]. While they can generally capture extreme emotions and recognize scenarios requiring urgent action, such as suicide ideation [54,65,70,72,78], they often fail to provide direct, practical measures, typically only advising users to seek professional help [96]. In addition, the inherent bias in LLM training data [66,106] can lead to the propagation of stereotypical, discriminatory, or biased viewpoints. This bias can also give rise to hallucinations, where LLMs produce erroneous or misleading information [105,131]. Hallucinations also may stem from overfitting the training data or a lack of context understanding [134]. Such inaccuracies can have serious consequences, such as providing incorrect medical information, reinforcing harmful stereotypes, or failing to recognize and appropriately respond to mental health crises [131]. For example, an LLM might reinforce a harmful belief held by a user, potentially exacerbating their mental health issues, or it could generate non-factual, overly optimistic, or pessimistic medical advice, delaying appropriate professional intervention. These issues could undermine the integrity and fairness of social psychology [102,105,106,110].

Another critical concern is the 'black box' nature of LLMs [105,107,131]. This lack of interpretability complicates the application of LLMs in mental health, where trustworthiness and clarity are important. When we talk about neural networks as black boxes, we know what they were trained with, how they were trained, what the weights are, etc. However, with many new LLMs like GPT-3.5/4, researchers and practitioners often access the models via web interfaces or APIs without complete knowledge of the training data, methods, and model updates. This situation not only presents the traditional challenges associated with neural networks but also has all these additional problems that come from the "hidden" model.

Ethical concern is another significant challenge associated with applying LLMs in mental health. Debates are emerging around issues like digital personhood, informed consent, the risk of manipulation, and the appropriateness of AI in mimicking human interactions [60,102,105,106,135]. A primary ethical concern is the potential alteration of the traditional therapist-patient relationship. Individuals may struggle to fully grasp the advantages and disadvantages of LLM derivatives, often choosing these options for their lower cost or greater convenience. This trend could lead to an increased reliance on the emotional support provided by AI [14], inadvertently positioning AI as the primary diagnostician and decision-maker for mental health issues, thereby undermining trust in conventional healthcare settings. Moreover, therapists may become overly reliant on LLM-generated answers and use them in clinical decision-making, overlooking the complexities involved in clinical assessment. This reliance could compromise their professional judgment and reduce opportunities for in-depth engagement with patients [17,129,130]. Furthermore, the dehumanization and technocratic nature of mental health care has the potential to depersonalize and dehumanize patients [136], where decisions are more driven by algorithms than by human insight and empathy. This can lead to decisions becoming mechanized, lacking empathy, and detached from ethics [137]. AI systems may fail to recognize or adequately interpret the subtle and often non-verbal cues critical in traditional therapeutic settings [136], such as tone of voice, facial expressions, and the emotional weight behind words, which are essential for comprehensively understanding a patient's condition and providing empathetic care.

Additionally, the current roles and accuracy of LLMs in mental health are limited. For instance, while LLMs can categorize a patient's mood or symptoms, most of these categorizations are binary, such as 'depressed' or 'not depressed' [50,65]. This oversimplification can lead to misdiagnoses. Data security and user privacy in clinical settings are also of utmost concern [14,54,60,96,130]. Although nearly 70% of psychiatrists believe that managing medical documents will be more efficient using LLMs, many still have concerns about their reliability and privacy [97,130,131]. These concerns could have a devastating impact on patient privacy and undermine the trust between physicians and patients if confidential treatment records stored in LLM databases are compromised. Beyond the technical limitations of AI, the current lack of an industry-benchmarked ethical framework and accountability system hinders the true application of LLMs in clinical practice [131].

### 4.2.2 Limitations of the Selected Articles

Several limitations were identified in the literature review. A significant issue is the age bias present in the social media data used for depression and mental health screening. Social media platforms tend to attract younger demographics, leading to an underrepresentation of older age groups [65]. Furthermore, most studies have focused on social media platforms primarily used by English-speaking populations, such as Twitter, which may result in a lack of insight into mental health trends in non-English-speaking regions. Our review included studies in Polish, Chinese, Portuguese, and Malay, all of which highlighted significant limitations of LLMs caused by the availability and size of databases [54,61,92,98,116]. For instance, due to the absence of a dedicated Polish-language mental health database, a Polish study had to rely on machine-translated English databases [92]. While the LLMs achieve 80% accuracy in categorizing emotions and moods in Polish, this is still lower than the 90% accuracy observed in the original English dataset. This discrepancy highlights that the accuracy of LLMs can be affected by the quality of the database.

Another limitation of our review is the low diversity of LLMs studied. Although we used 'large language models' as keywords in our search phase, most identified studies primarily focused on BERT and its variants, as well as GPT models. Therefore, this review provides only a limited picture of the variability we might expect in applicability between different LLMs. Additionally, the rapid development of LLM technologies presents a limitation; our study can only reflect current trends and may not encompass future advances or the full potential of LLMs. For instance, in tests involving psychologically relevant questions and answers, ChatGPT 3.5 achieved an accuracy of 66.8%, while ChatGPT 4.0 reached an accuracy of 85%, compared to an average human score of 73.8% [118]. Evaluating ChatGPT at different stages separately and comparing its performance to that of humans can lead to varied conclusions. In the assessment of prognosis and treatment planning for depression using LLMs, ChatGPT 3.5 demonstrated a distinctly pessimistic prognosis that differed significantly from those of ChatGPT-4, Claude, Bard, and mental health professionals [128]. Therefore, continuous monitoring and evaluation are essential to fully understand and effectively utilize the advancements in LLM technologies.

### 4.3 Opportunities and Future Work

Implementing technologies involving LLMs within the healthcare provision of real patients demands thorough and multi-faceted evaluations. It is imperative for both industry and researchers to not let rollout exceed proportional requirements for evidence on safety and efficacy. At the level of the service provider, this includes providing explicit warnings to the public to discourage mistaking LLM functionality for clinical reliability. For example, ChatGPT-4 introduced the ability to process and interpret image inputs within conversational contexts, leading OpenAI to issue an official warning that ChatGPT-4 is not approved for analyzing specialized medical images, such as CT scans [138].

A key challenge to address in LLM research is the tendency to produce incoherent text or hallucinations. Future efforts could focus on training LLMs specifically for mental health applications, using datasets with expert labeling to reduce bias and create specialized mental health lexicons [84,102,116]. The creation of specialized datasets could take advantage of the customizable nature of LLMs, fostering the development of models that cater to the distinct needs of varied demographic groups. For instance, unlike models designed for healthcare professionals which assist in tasks like data documentation, symptom analysis, medication management, and postoperative care, LLMs intended for patient interaction might be trained with an emphasis on empathy and comfortable dialogue.

Another critical concern is the problem of outdated training data in LLMs. Traditional LLMs, such as GPT-4 (with a cut-off in October 2023), rely on potentially outdated training data, limiting their ability to incorporate recent events or information. This can compromise the accuracy and relevance of their responses, leading to the generation of uninformative or incorrect answers, known as 'hallucinations' [139]. RAG (Retrieval-Augmented Generation) technology offers a solution by retrieving facts from external knowledge bases, ensuring that LLMs use the most accurate and up-to-date information [140]. By searching for relevant information from numerous documents, RAG enhances the generation process with the most recent and contextually relevant content [141]. Additionally, RAG includes evidence-based information, increasing the reliability and credibility of LLM responses [139].

To further enhance the reliability of LLM content and minimize hallucinations, recent studies suggest adjusting model parameters, such as the 'temperature' setting [142-144]. The 'temperature' parameter influences the randomness and predictability of outputs [145]. Lowering the temperature typically results in more deterministic outputs, enhancing coherence and reducing irrelevant content [146]. However, this adjustment can also limit the model's creativity and adaptability, potentially making it less effective in scenarios requiring diverse or nuanced responses. In mental therapy, where nuanced and sensitive responses are essential, maintaining an optimal balance is crucial. While a lower temperature can ensure accuracy, which is important for tasks like clinical documentation, it may not suit therapeutic dialogs where personalized engagement is key. Low temperatures can lead to repetitive and impersonal responses, reducing patient engagement and therapeutic effectiveness. To mitigate these risks, regular updates of the model incorporating the latest therapeutic practices and clinical feedback are essential. Such updates could refine the model's understanding and response mechanisms, ensuring it remains a safe and effective tool for mental health care. Nevertheless, determining the 'optimal' temperature setting is challenging, primarily due to the variability in tasks and interaction contexts which require different levels of creativity and precision.

Data privacy is another important area of concern. Many LLMs, such as ChatGPT and Claude, involve sending data to third-party servers, which poses the risk of data leakage. Current studies have found that LLMs can be enhanced by Privacy Enhancing Techniques, such as zero-knowledge proofs, differential privacy, and federated learning [147]. Additionally, privacy can be preserved by replacing identifying information in textual data with generic tokens. For example, when recording sensitive information (e.g., names, addresses, or credit card numbers), using alternatives to mask tokens can help protect user data from unauthorized access [148]. This obfuscation technique ensures that sensitive user information is not stored directly, thereby enhancing data security.

The lack of interpretability in LLM decision-making is another crucial area for future research on healthcare applications. Future research should examine the models' architecture, training, and inferential processes for clearer understanding. Detailed documentation of training datasets, sharing of model architectures, and third-party audits would ideally form part of this undertaking. Investigating techniques like attention mechanisms and modular architectures could illuminate aspects of neural network processing. The implementation of knowledge graphs might help in outlining logical relationships and facts [149]. Additionally, another promising approach involves creating a dedicated embedding space during training, guided by an LLM. This space aligns with a causal graph and aids in identifying matches that approximate counterfactuals [150].

Before deploying LLMs in mental health settings, a comprehensive assessment of their reliability, safety, fairness, abuse resistance, interpretability, compliance with social norms, robustness, performance, linguistic accuracy, and cognitive ability is essential. It is also crucial to foster collaborative relationships among mental health professionals, patients, AI researchers, and policymakers. LLMs, for instance, have demonstrated initial competence in providing medication advice, yet their responses can sometimes be inconsistent or include inappropriate suggestions. As such, LLMs require professional oversight and should not be used independently. However, when utilized as decision aids, LLMs have the potential to enhance healthcare efficiency. We call on developers of LLMs to collaborate with authoritative regulators in actively developing ethical guidelines for AI research in healthcare. These guidelines should aim to adopt a balanced approach that considers the multifaceted nature of LLMs and ensures their responsible integration into medical practice. They are expected to become industry benchmarks, facilitating the future development of LLMs in mental health.

### 4.4 Conclusion

This review examines the use of LLMs in mental health applications, including text-based screening for mental health conditions, detection of suicidal ideation, CAs, clinical use, and other related applications. Despite their potential, challenges such as the production of hallucinatory or harmful information, output inconsistency, and ethical concerns remain. Nevertheless, as technology advances and ethical guidelines improve, LLMs are expected to become increasingly integral and valuable in mental health services, providing alternative solutions to this global healthcare issue.

### 4.5 Contributors

ZG and KL contributed to the conception and design of the study. ZG, KL, and AL also contributed to the development of the search strategy. Database search outputs were screened by ZG, and data were extracted by ZG and KL. An assessment of the risk of bias in the included studies was performed by ZG and KL. ZG completed the literature review, collated the data, performed the data analysis, interpreted the results, and wrote the first draft of the manuscript. KL, AL, JHT, JF, and TK reviewed the manuscript and provided multiple rounds of guidance in the writing of the manuscript. All authors read and approved the final version of the manuscript.

### 4.6 Acknowledgements

This work was funded by the UKRI Centre for Doctoral Training in AI-enabled healthcare systems (grant EP/S021612/1). The funders were not involved in the study design, data collection, analysis, publication decisions, or manuscript writing. The views expressed in the text are those of the authors and not those of the funder.

### 4.7 Conflicts of Interest

The authors declare no conflict of interest.

### 4.8 Data sharing statement

The authors ensure that all pertinent data have been incorporated within the article and/or its supplementary materials. For access to the research data, interested parties may contact the corresponding author, Kezhi Li (ken.li@ucl.ac.uk), subject to a reasonable request.

### 4.9 Abbreviations

**ACM:** ACM Digital Library
**AI:** Artificial Intelligence
**ADHD:** Attention Deficit Hyperactivity Disorder
**BERT:** Bidirectional Encoder Representations from Transformers
**BPD:** Borderline Personality Disorder
**CNN:** Convolutional Neural Network
**CA:** Conversational Agent
**GANs:** General Adversarial Networks
**GPT:** Generative Pre-trained Transformer
**JMIR:** Journal of Medical Internet Research
**KL:** Kuala Lumpur
**LEAS:** Levels of Emotional Awareness Scale
**LLM:** Large Language Model
**LSTM:** Long Short-Term Memory Networks
**ML:** Machine Learning
**MAE:** Mean Absolute Error
**NLP:** Natural Language Processing
**PPD:** Postpartum Depression
**PRISMA:** Preferred Reporting Items for Systematic Review and Meta-Analysis
**SCPQ:** Stress And Coping Process Questionnaire
**SPD:** Schizoid Personality Disorder
**WHO:** World Health Organization

# 5. Multimedia Appendix 1

## Supplementary material 1: Risk of bias assessment

| Study | D1 | D2 | D3 | D4 | D5 | Overall |
|---|---|---|---|---|---|---|
| Verma et al., 2023 | + | + | + | + | + | + |
| Diniz et al., 2022 | + | + | + | + | + | + |
| Danner et al., 2023 | + | + | − | + | + | − |
| Tao et al., 2023 | + | + | + | + | + | + |
| Hayati et al., 2022 | + | + | − | + | + | − |
| Wang et al., 2020 | + | + | − | + | + | − |
| Metzler et al., 2022 | + | + | + | + | + | + |
| Sadeghi et al., 202 | + | + | − | + | + | − |
| Zhang et al., 2021 | − | + | − | + | + | − |
| Vajre et al., 2021 | + | + | + | + | + | + |
| Levkovich & Elyoseph, 2023 | + | + | + | + | + | + |
| Howard et al., 2020 | + | + | + | + | + | + |
| Beredo & Ong, 2022 | + | + | + | + | + | + |
| Crasto et al., 2021 | + | + | + | + | + | + |
| Zygadlo, 2021 | + | × | × | + | + | × |
| Ma et al., 2023 | + | + | + | + | + | + |
| Heston, 2023 | + | + | + | + | + | + |
| Franco D'Souza et al., 2023 | + | + | − | + | + | − |
| Spallek et al., 2023 | + | + | + | + | + | + |
| Farhat et al., 2023 | + | + | + | + | + | + |
| Wei et al., 2023 | + | + | + | + | + | + |
| Yongsatianchot et al., 2023 | + | + | + | + | + | + |
| Grabb, 2023 | − | + | + | + | + | − |
| Hadar-Shoval et al., 202 | + | + | + | + | + | + |
| Sezgin et al., 2023 | + | + | + | + | + | + |
| Tanana et al., 2021 | + | + | + | + | + | + |
| Wang et al., 2023 | − | + | + | − | + | − |
| Schubert et al., 2023 | − | + | + | − | + | − |
| (Friedman & Ballentine, 2023) | − | + | − | + | + | + |
| (Wu et al., 2023) | + | + | + | + | + | + |
| (Stigall et al., 2024) | + | + | − | + | + | + |
| (Ghanadian et al., 2024) | + | + | − | + | + | + |
| (Lossio-Ventura et al., 2024) | + | + | + | + | + | + |
| (Alessa and Al-Khalifa, 2023) | − | + | − | + | − | − |
| (He et al., 2024) | − | + | + | + | + | + |
| (Kumar et al., 2023) | + | + | + | + | + | + |
| (Elyoseph et al., 2024) | + | + | + | + | + | + |
| (Perlis et al., 2024) | + | + | − | + | − | − |
| (Blease et al., 2024) | − | + | − | + | + | − |
| (Berrezueta-Guzman et al., 2024) Evaluation | + | + | + | + | + | + |

Domains:
D1: Bias arising from the randomization process.
D2: Bias due to deviations from intended intervention.
D3: Bias due to missing outcome data.
D4: Bias in measurement of the outcome.
D5: Bias in selection of the reported result.

Judgement:
● High
● Some concerns
● Low

Table S1: Risk of bias assessment

# 6. Multimedia Appendix 2

**Supplementary material 2: List of studies excluded at the full-text screening stage**

| | Title | Reference | Exclusion reason |
|---|---|---|---|
| 1 | A Novel AI-based chatbot Application for Personalized Medical Diagnosis and review using Large Language Models | (S et al., 2023) | Not about mental health |
| 2 | An Introduction to Generative Artificial Intelligence in Mental Health Care: Considerations and Guidance | (King et al., 2023) | Not only for mental health |
| 3 | Artificial Intelligence Based Analysis of Positive and Negative Tweets Towards COVID-19 Vaccines | (Umair & Masciari, 2021) | Not about mental health |
| 4 | Artificial Intelligence in Psychiatry | (Briganti, 2023) | Not about LLMs |
| 5 | Assessing the Accuracy of Responses by the Language Model ChatGPT to Questions Regarding Bariatric Surgery | (Samaan et al., 2023) | Not about mental health |
| 6 | ChatGPT and Bard Exhibit Spontaneous Citation Fabrication during Psychiatry Literature Search | (McGowan et al., 2023) | Not about mental health |
| 7 | ChatGPT and mental healthcare: balancing benefits with risks of harms | (Blease & Torous, 2023) | It's a review paper, too short |
| 8 | ChatGPT in Answering Queries Related to Lifestyle-Related Diseases and Disorders | (Mondal et al., 2023) | Not about mental health |
| 9 | ChatGPT on ECT: Can Large Language Models Support Psychoeducation? | (Lundin et al., 2023) | It's a letter, It's too short |
| 10 | ChatGPT vs Google for Queries Related to Dementia and Other Cognitive Decline: Comparison of Results | (Hristidis et al., 2023) | Dementia is not considered a mental illness |
| 11 | ChatGPT vs. Human Annotators: A Comprehensive Analysis of ChatGPT for Text Annotation | (Aldeen et al., 2023) | Not about mental health |
| 12 | Conversational Agents in Health Care: Expert Interviews to Inform the Definition, Classification, and Conceptual Framework | (Martinengo et al., 2023) | Not about LLMs |
| 13 | Diagnosing Psychiatric Disorders from History of Present Illness Using a Large-Scale Linguistic Model | (Otsuka et al., 2023) | Not about LLMs |
| 14 | Discourse-Level Representations Can Improve Prediction of Degree of Anxiety | (Juhng et al., 2023) | Duplicate |
| 15 | Empathy and Equity: Key Considerations for Large Language Model Adoption in Health Care | (Koranteng et al., 2023) | Not about mental health |
| 16 | Ethical Challenges in AI Approaches to Eating Disorders | (Sharp et al., 2023) | Not about LLMs |
| 17 | Evaluating the Application of Large Language Models in Clinical Research Contexts | (Perlis & Fihn, 2023) | It's a review paper, too short |
| 18 | Examining the Utility of Social Media in COVID-19 Vaccination: Unsupervised Learning of 672,133 Twitter Posts | (Liew & Lee, 2021) | Not about LLMs |
| 19 | Global Mental Health Services and the Impact of Artificial Intelligence–Powered Large Language Models | (van Heerden et al., 2023) | It's a review paper, too short |
| 20 | Grateful Chatbots: Public Sensemaking through Individual Gratitude Interventions | (Schuler & Portmann, 2023) | Not about mental health |
| 21 | Identifying Rare Circumstances Preceding Female Firearm Suicides: Validating A Large Language Model Approach | (Zhou et al., 2023) | Not about mental health |
| 22 | The Impact of Multimodal Large Language Models on Health Care's Future | (Meskó, 2023) | Not about mental health |
| 23 | Large Language Models in Medical Education: Opportunities, Challenges, and Future Directions | (Abd-alrazaq et al., 2023) | Not about mental health |
| 24 | Linguistic Features of Clients and Counselors for Early Detection of Mental Health Issues in Online Text-based Counseling | (Shidara et al., 2022) | Not about LLMs |
| 25 | Mental Health Prediction from Social Media Text Using Mixture of Experts | (Santos et al., 2023) | Duplicate |
| 26 | Negatively Correlated Noisy Learners for At-Risk User Detection on Social Networks: A Study on Depression, Anorexia, Self-Harm, and Suicide | (Ragheb et al., 2023) | Duplicate |
| 27 | Performance of ChatGPT on the Situational Judgement Test—A Professional Dilemmas–Based Examination for Doctors in the United Kingdom | (Borchert et al., 2023) | Not about mental health |
| 28 | Predicting Generalized Anxiety Disorder from Impromptu Speech Transcripts Using Context-Aware Transformer-Based Neural Networks: Model Evaluation Stud | (Teferra & Rose, 2023) | Not about LLMs |
| 29 | Psychological Insights into The Research and Practice of Embodied Conversational Agents, Chatbots and Social Assistive Robots: A Systematic Meta-Review | (Kiuchi et al., 2023) | Not about LLMs |
| 30 | Automatic rating of therapist facilitative interpersonal skills in text: A natural language processing application | (Zech et al., 2022) | Not about mental health |
| 31 | Social Media Images Can Predict Suicide Risk Using Interpretable Large Language-Vision Models | (Badian et al., 2023) | Duplicate |
| 32 | Systematic review and meta-analysis of AI-based conversational agents for promoting mental health and well-being | (Li et al., 2023) | Not about LLMs |
| 33 | Text Dialogue Analysis for Primary Screening of Mild Cognitive Impairment: Development and Validation Study | (C. Wang et al., 2023) | Not about mental health |
| 34 | The Impact of Multimodal Large Language Models on Health Care's Future | (Meskó, 2023) | Not about mental health |
| 35 | Transformer-based deep neural network language models for Alzheimer's disease risk assessment from targeted speech | (Roshanzamir et al., 2021) | Not about mental health |
| 36 | Understanding Dyslexia Through Personalized Large-Scale Computational Models | (Perry et al., 2019) | Not about mental health |
| 37 | Using Generative Artificial Intelligence to Classify Primary Progressive Aphasia from Connected Speech | (Rezaii et al., 2023) | It's preprint |
| 38 | Waiting for A Digital Therapist: Three Challenges on the Path to Psychotherapy Delivered by Artificial Intelligence | (Grodniewicz & Hohol, 2023) | Not about LLMs |
| 39 | A Transfer Learning Method for Detecting Alzheimer's Disease Based on Speech and Natural Language Processing | (Liu et al., 2022) | Not about mental health |
| 40 | Acoustic and Linguistic Analyses to Assess Early-Onset and Genetic Alzheimer's Disease | (Pérez-Toro et al., 2021) | Not about mental health |
| 41 | Using a Chatbot to Provide Formative Feedback: A Longitudinal Study of Intrinsic Motivation, Cognitive Load, and Learning Performance | (Yin et al., 2024) | Duplicate |
| 42 | Leveraging Large Language Models for Improved Patient Access and Self-Management: Assessor-Blinded Comparison Between Expert- and AI-Generated Content | (Lv et al., 2024) | Not about mental health |
| 43 | Evaluation of Prompts to Simplify Cardiovascular Disease Information Generated Using a Large Language Model: Cross-Sectional Study | (Mishra et al., 2024) | Not about mental health |
| 44 | Evaluation of the Performance of Generative AI Large Language Models ChatGPT, Google Bard, and Microsoft Bing Chat in Supporting Evidence-Based Dentistry: Comparative Mixed Methods Study | (Giannakopoulos et al., 2023) | Not about mental health |
| 45 | Beyond Discrimination: Generative AI Applications and Ethical Challenges in Forensic Psychiatry | (Tortora, 2024) | Not about LLMs |
| 46 | Assessing the Alignment of Large Language Models With Human Values for Mental Health Integration: Cross-Sectional Study Using Schwartz's Theory of Basic Values | (Hadar-Shoval et al., 2024) | Not about mental health |
| 47 | Depression and Reciprocal Language Style Matching in Text Messages | (Weinstein and Jensen, 2024) | Not about LLMs |
| 48 | "I Have a Different Perspective as I Am Working Through This" Speech–Language Pathologist Reflections on Autism | (DeThorne et al., 2024) | Not about LLMs |
| 49 | Artificial Intelligence in Medical Education: Comparative Analysis of ChatGPT, Bing, and Medical Students in Germany | (Roos et al., 2023) | Not about mental health |
| 50 | An Entity Extraction Pipeline for Medical Text Records Using Large Language Models: Analytical Study | (Wang et al., 2024) | Not about mental health |
| 51 | XAI Transformer based Approach for Interpreting Depressed and Suicidal User Behavior on Online Social Networks | (Malhotra and Jindal, 2024) | Duplicate |
| 52 | University Students' Acceptance and Usage of Generative AI (ChatGPT) from a Psycho-Technical Perspective | (Faruk et al., 2023) | Not about mental health |
| 53 | Fairness Evaluation Within Large Language Models through the Lens of Depression | (Han, 2024) | Too short |
| 54 | A Machine Learning Enabled Approach for Mental and Physical Health Management Using OpenCV, NLP and IOT | (Rane et al., 2024) | Not about LLMs |
| 55 | Machine Feeling by Knowledge Acquisition with Emotion Map | (Lim et al., 2024) | Not about LLMs |
| 56 | Evaluating Emotional Detection & Classification Capabilities of GPT-2 & GPT-Neo Using Textual Data | (Jain et al., 2024) | Duplicate |
| 57 | Development of Serious Game Theory Framework in Virtual Reality for Alzheimer's Patients | (Zuo et al., 2024) | Not about LLMs |
| 58 | Calibration of Transformer-Based Models for Identifying Stress and Depression in Social Media | (Ilias et al., 2023) | Duplicate |
| 59 | ALTRUIST: a Python package to emulate a Virtual Digital Cohort Study using social media data | (Bour et al., 2024) | Not about LLMs |
| 60 | Large Language Models and Healthcare Alliance: Potential and Challenges of Two Representative Use Cases | (García-Méndez and de Arriba-Pérez, 2024) | Not about mental health |
| 61 | A Platform for Connecting Social Media Data to Domain-Specific Topics Using Large Language Models: An Application to Student Mental Health | (Ruocco et al., 2024) | Duplicate |

Table S2: Studies excluded after full-text screening. LLMs=large language models

# 7. Multimedia Appendix 3

## Supplementary material 3: PRISMA Checklist

| Section and Topic | Item # | Checklist item | Location where item is reported |
|---|---|---|---|
| **TITLE** | | | |
| Title | 1 | Identify the report as a systematic review. | Title Page- Pg 1 |
| **ABSTRACT** | | | |
| Abstract | 2 | See the PRISMA 2020 for Abstracts checklist. | Abstract- Pg 1-2 |
| **INTRODUCTION** | | | |
| Rationale | 3 | Describe the rationale for the review in the context of existing knowledge. | Introduction- Pg 2-6 |
| Objectives | 4 | Provide an explicit statement of the objective(s) or question(s) the review addresses. | Introduction- Pg 6 |
| **METHODS** | | | |
| Eligibility criteria | 5 | Specify the inclusion and exclusion criteria for the review and how studies were grouped for the syntheses. | Methods- Pg 7-8 |
| Information sources | 6 | Specify all databases, registers, websites, organisations, reference lists and other sources searched or consulted to identify studies. Specify the date when each source was last searched or consulted. | Methods- Pg 7-8 |
| Search strategy | 7 | Present the full search strategies for all databases, registers and websites, including any filters and limits used. | Methods- Pg 6-7 |
| Selection process | 8 | Specify the methods used to decide whether a study met the inclusion criteria of the review, including how many reviewers screened each record and each report retrieved, whether they worked independently, and if applicable, details of automation tools used in the process. | Methods- Pg 7-8 |
| Data collection process | 9 | Specify the methods used to collect data from reports, including how many reviewers collected data from each report, whether they worked independently, any processes for obtaining or confirming data from study investigators, and if applicable, details of automation tools used in the process. | Methods- Pg 8-9 |
| Data items | 10a | List and define all outcomes for which data were sought. Specify whether all results that were compatible with each outcome domain in each study were sought (e.g. for all measures, time points, analyses), and if not, the methods used to decide which results to collect. | Methods- Pg 8 |
| | 10b | List and define all other variables for which data were sought (e.g. participant and intervention characteristics, funding sources). Describe any assumptions made about any missing or unclear information. | Methods- Pg 8 |
| Study risk of bias assessment | 11 | Specify the methods used to assess risk of bias in the included studies, including details of the tool(s) used, how many reviewers assessed each study and whether they worked independently, and if applicable, details of automation tools used in the process. | Methods- Pg 7; Multimedia Appendix 1 |
| Effect measures | 12 | Specify for each outcome the effect measure(s) (e.g. risk ratio, mean difference) used in the synthesis or presentation of results. | Methods- Pg 8 |
| Synthesis methods | 13a | Describe the processes used to decide which studies were eligible for each synthesis (e.g. tabulating the study intervention characteristics and comparing against the planned groups for each synthesis (item #5)). | Methods- Pg 7-8 |
| | 13b | Describe any methods required to prepare the data for presentation or synthesis, such as handling of missing summary statistics, or data conversions. | Methods- Pg 8 |
| | 13c | Describe any methods used to tabulate or visually display results of individual studies and syntheses. | Methods- Pg 7-8 |
| | 13d | Describe any methods used to synthesize results and provide a rationale for the choice(s). If meta-analysis was performed, describe the model(s), method(s) to identify the presence and extent of statistical heterogeneity, and software package(s) used. | Methods- Pg 7-8 |
| | 13e | Describe any methods used to explore possible causes of heterogeneity among study results (e.g. subgroup analysis, meta-regression). | Methods- Pg 8 |
| | 13f | Describe any sensitivity analyses conducted to assess robustness of the synthesized results. | |
| Reporting bias assessment | 14 | Describe any methods used to assess risk of bias due to missing results in a synthesis (arising from reporting biases). | Methods- Pg 7 |
| Certainty assessment | 15 | Describe any methods used to assess certainty (or confidence) in the body of evidence for an outcome. | Methods- Pg 7-8 |
| **RESULTS** | | | |
| Study selection | 16a | Describe the results of the search and selection process, from the number of records identified in the search to the number of studies included in the review, ideally using a flow diagram. | Results- Pg 9-10 |

| Section and Topic | Item # | Checklist item | Location where item is reported |
|---|---|---|---|
| | 16b | Cite studies that might appear to meet the inclusion criteria, but which were excluded, and explain why they were excluded. | Results- Pg 9; Multimedia Appendix 2 |
| Study characteristics | 17 | Cite each included study and present its characteristics. | Results- Pg 11-15 (Table1-3) |
| Risk of bias in studies | 18 | Present assessments of risk of bias for each included study. | Multimedia Appendix 1 |
| Results of individual studies | 19 | For all outcomes, present, for each study: (a) summary statistics for each group (where appropriate) and (b) an effect estimate and its precision (e.g. confidence/credible interval), ideally using structured tables or plots. | Results- Pg 11-17 (Table1-3) |
| Results of syntheses | 20a | For each synthesis, briefly summarise the characteristics and risk of bias among contributing studies. | Results- Pg 18-21 |
| | 20b | Present results of all statistical syntheses conducted. If meta-analysis was done, present for each the summary estimate and its precision (e.g. confidence/credible interval) and measures of statistical heterogeneity. If comparing groups, describe the direction of the effect. | Results - Pg 18-21 |
| | 20c | Present results of all investigations of possible causes of heterogeneity among study results. | Results - Pg 18-21 |
| | 20d | Present results of all sensitivity analyses conducted to assess the robustness of the synthesized results. | |
| Reporting biases | 21 | Present assessments of risk of bias due to missing results (arising from reporting biases) for each synthesis assessed. | Results - Pg 18-22; Multimedia Appendix 1 |
| Certainty of evidence | 22 | Present assessments of certainty (or confidence) in the body of evidence for each outcome assessed. | Results - Pg 11-22 |
| **DISCUSSION** | | | |
| Discussion | 23a | Provide a general interpretation of the results in the context of other evidence. | Discussion- Pg 24 |
| | 23b | Discuss any limitations of the evidence included in the review. | Discussion- Pg 24-26 |
| | 23c | Discuss any limitations of the review processes used. | Discussion- Pg 26-27 |
| | 23d | Discuss implications of the results for practice, policy, and future research. | Discussion- Pg 27-29 |
| **OTHER INFORMATION** | | | |
| Registration and protocol | 24a | Provide registration information for the review, including register name and registration number, or state that the review was not registered. | Methods- Pg 6-7 |
| | 24b | Indicate where the review protocol can be accessed, or state that a protocol was not prepared. | Methods- Pg 6-7 |
| | 24c | Describe and explain any amendments to information provided at registration or in the protocol. | Methods- Pg 6-7 |
| Support | 25 | Describe sources of financial or non-financial support for the review, and the role of the funders or sponsors in the review. | Acknowledgment- Pg 30 |
| Competing interests | 26 | Declare any competing interests of review authors. | Conflicts of Interest- Pg 30 |
| Availability of data, code and other materials | 27 | Report which of the following are publicly available and where they can be found: template data collection forms; data extracted from included studies; data used for all analyses; analytic code; any other materials used in the review. | Data sharing statement- Pg 30 |

Table S3: PRISMA Checklist